\documentclass[sigconf]{acmart}
\AtBeginDocument{%
  \providecommand\BibTeX{{%
    \normalfont B\kern-0.5em{\scshape i\kern-0.25em b}\kern-0.8em\TeX}}}

\makeatletter
\def\@ACM@copyright@check@cc{}
\makeatother
\setcopyright{cc}

\copyrightyear{2025} 
\acmYear{2025}
\setcctype{by-nc-nd}
\acmConference[FAccT '25]{The 2025 ACM Conference on Fairness, Accountability, and Transparency}{June 23--26, 2025}{Athens, Greece}
\acmBooktitle{The 2025 ACM Conference on Fairness, Accountability, and Transparency (FAccT '25), June 23--26, 2025, Athens, Greece}\acmDOI{10.1145/3715275.3732155}
\acmISBN{979-8-4007-1482-5/2025/06}

\usepackage{graphicx}
\usepackage{multirow}
\usepackage{caption,subcaption,graphicx}
\usepackage{xcolor}

\begin{document}

\title{Towards Uncertainty Aware Task Delegation and Human-AI Collaborative Decision-Making}


\author{Min Hun Lee}
\email{mhlee@smu.edu.sg}
\affiliation{%
  \institution{Singapore Management University}
  \city{Singapore}
  \country{Singapore}
}

\author{Martyn Zhe Yu Tok}
\orcid{1234-5678-9012}
\email{martyn.tok.2022@scis.smu.edu.sg}
\affiliation{%
  \institution{Singapore Management University}
  \city{Singapore}
  \country{Singapore}
}

\renewcommand{\shortauthors}{Lee and Tok}

\begin{abstract}
Despite the growing promise of artificial intelligence (AI) in supporting decision-making across domains, fostering appropriate human reliance on AI remains a critical challenge. In this paper, we investigate the utility of exploring distance-based uncertainty scores for task delegation to AI and describe how these scores can be visualized through embedding representations for human-AI decision-making. After developing an AI-based system for physical stroke rehabilitation assessment, we conducted a study with 19 health professionals and 10 students in medicine/health to understand the effect of exploring distance-based uncertainty scores on users' reliance on AI. Our findings showed that distance-based uncertainty scores outperformed traditional probability-based uncertainty scores in identifying uncertain cases. In addition, after exploring confidence scores for task delegation and reviewing embedding-based visualizations of distance-based uncertainty scores, participants achieved an 8.20\% higher rate of correct decisions, a 7.15\% higher rate of changing their decisions to correct ones, and a 7.14\% lower rate of incorrect changes after reviewing AI outputs than those reviewing probability-based uncertainty scores ($p<0.01$). Our findings highlight the potential of distance-based uncertainty scores to enhance decision accuracy and appropriate reliance on AI while discussing ongoing challenges for human-AI collaborative decision-making.
\end{abstract}






\maketitle

\section{Introduction}
Advanced artificial intelligence (AI) is increasingly being used to develop decision support systems that provide data-driven insights to improve various decision-making tasks (e.g., health \cite{beede2020human,cai2019human,lee2021human,wang2021brilliant,caruana2015intelligible} and other social services \cite{kuo2023understanding,zavrvsnik2020criminal}). 
Instead of relying solely on fully autonomous systems, a human-AI collaborative approach is a prominent paradigm that investigates how humans and AI systems can complement each other's strengths \cite{beede2020human,nam2019development,singh2018deep,lee2021human,lai2021towards,cai2021onboarding} and surpass the capabilities of either human or AI alone and achieve synergistic, complementary performance \cite{cai2019human,topol2019high,lee2021human}. 

For human-AI collaborative approaches, researchers have investigated computational techniques to make AI more explainable \cite{arya2019one,wang2019designing,lakkaraju2020explaining,abdul2018trends,preece2018asking} and to communicate the uncertainty of AI outputs \cite{smith2013uncertainty,hendrycks2016baseline,corbiere2019addressing,gal2016theoretically,kendall2017uncertainties,jiang2018trust,papernot2018deep,van2020uncertainty}. Providing AI explanations or uncertainty information could enhance users' understanding of AI models and improve their decision-making. However, prior work also highlights challenges that presenting AI explanations may lead to over-reliance on incorrect AI outputs \cite{bussone2015role,buccinca2021trust,lee2023understanding,chen2023understanding}. In addition, presenting uncertainty information does not necessarily guarantee to improve users' decision-making \cite{zhang2024evaluating}. While several studies have explored the effectiveness of presenting confidence scores during the AI-assisted decision-making  \cite{zhang2020effect,prabhudesai2023understanding,zhang2024evaluating}, there is limited understanding of how to encourage more analytical engagement with AI outputs including confidence scores \cite{zhang2020effect,prabhudesai2023understanding} and of the potential utility of uncertainty information prior to the AI-assisted decision-making phase for human-AI collaborative decision-making.

\begin{figure*}[htp]
\centering 
\begin{subfigure}[t]{0.34\textwidth}
  \centering
  \includegraphics[width=1.0\columnwidth]%{figures/Interface_ExploreConfid.png}
  {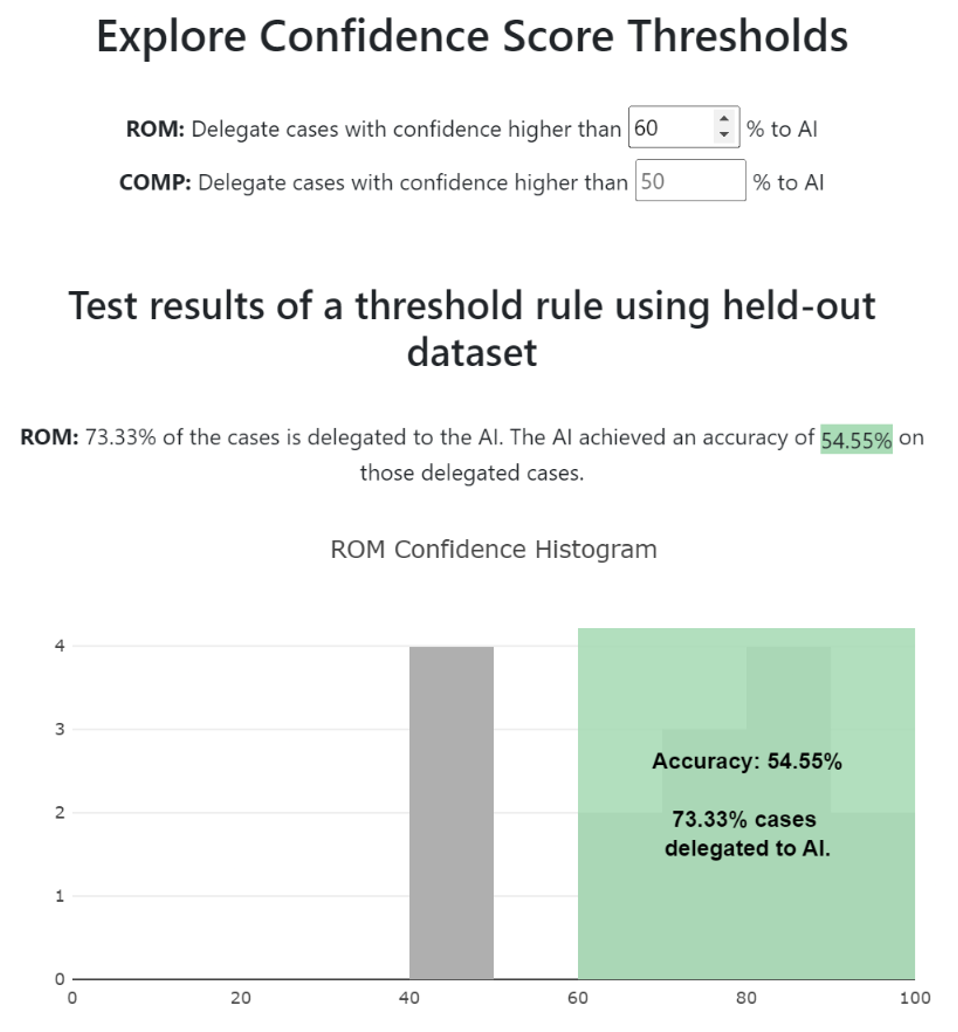}
  \caption{}
  \label{fig:interface_expconf}
\end{subfigure}
\begin{subfigure}[t]{0.53\textwidth}
  \centering
  \includegraphics[width=1.0\columnwidth]{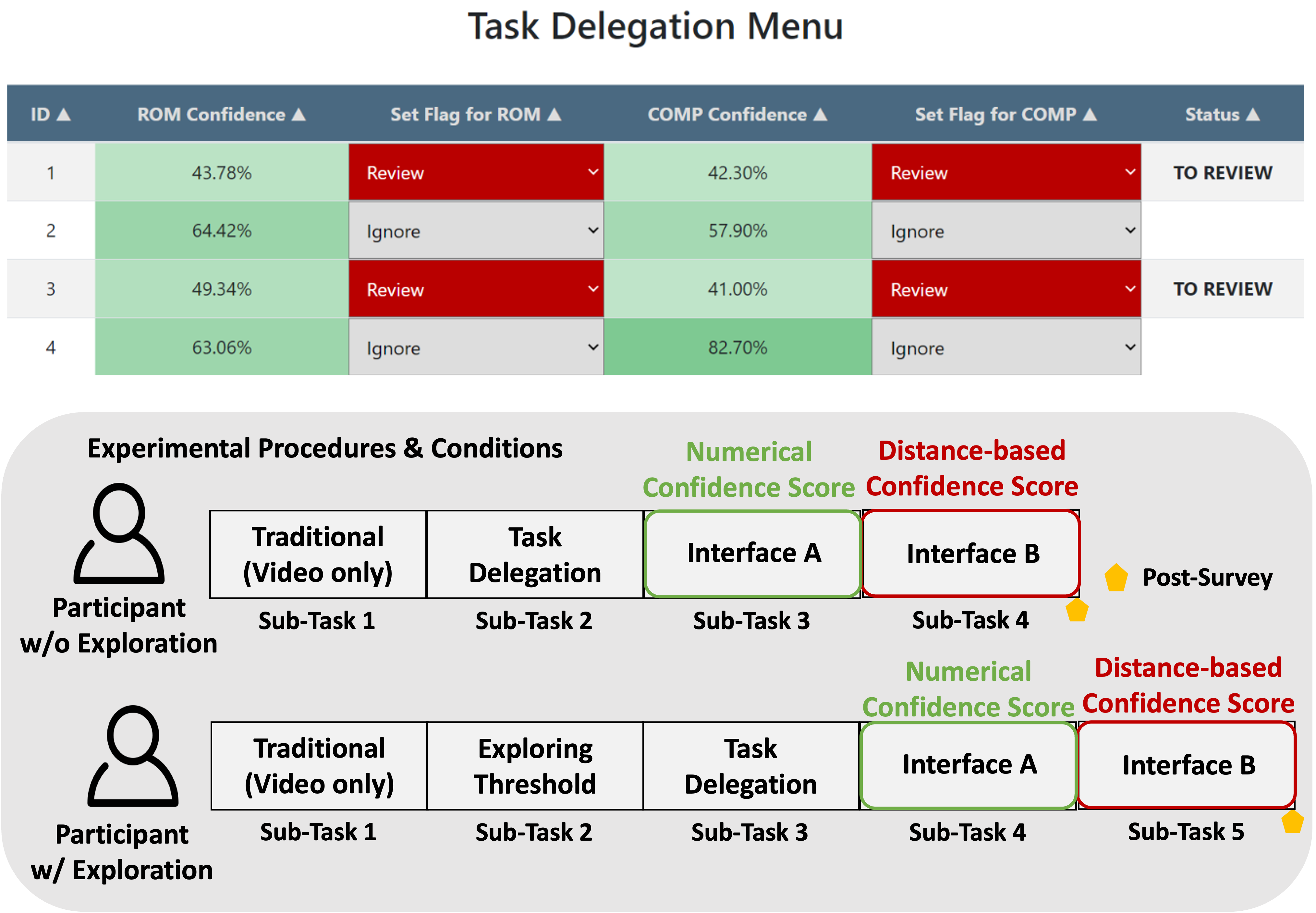}
  \caption{}
  \label{fig:interface_taskdelegate}
\end{subfigure}
\caption{(a) Users can explore different thresholds of confidence scores, review AI performance on delegated cases from a held-out dataset, and specify a confidence threshold to delegate cases to AI (e.g. delegating cases with confidence scores above 60\% to AI). (b) Task Delegation: Users can review AI confidence scores on assigned cases and identify which ones require human review. 2x2 Experimental Conditions: Participants with/without exploring a threshold interacted with two interfaces with numerical and distance-based confidence scores.}\label{fig:interface_before}
\end{figure*}

\begin{figure*}[htp]
\centering 
\begin{subfigure}[t]{0.75\textwidth}
  \centering
  \includegraphics[width=1.0\columnwidth]{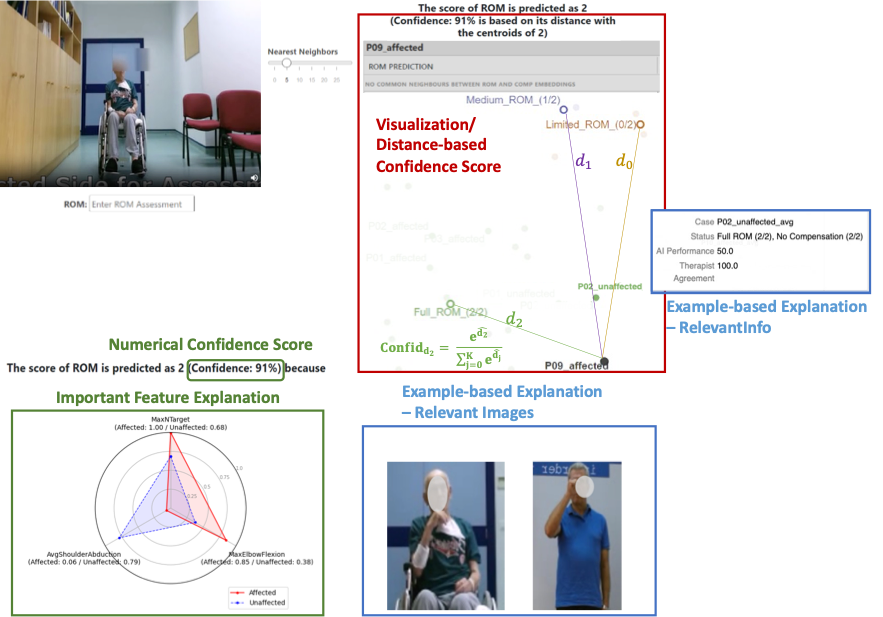}
\end{subfigure}
\caption{Interface for AI-assisted decision-making. For each case, the system shows the video of a patient along with an AI predicted score on patient's quality of motion, confidence scores, example-based explanations (i.e. relevant images and information), and important feature explanations. For confidence scores, our work investigates comparing a distance-based visualization of confidence scores with a numerical presentation of confidence scores. (1) Numerical - representing the highest class probability output by the model and (2) Distance-based confidence scores computed by measuring the distance between the input and the centroid of each class (among $K$ classes), normalizing it by (1 - $\frac{d}{d_{max}}$) and applying a softmax function.}\label{fig:interface_during}
\end{figure*}

In this work, we focus on the domain of physical stroke rehabilitation assessment and contribute to a study exploring how interactive exploration of AI confidence scores affects task delegation and user reliance during human-AI collaborative decision-making. Specifically, we investigate two key features: (1) interactive threshold-based exploration of confidence scores for task delegation (Figure \ref{fig:interface_before}) and (2) distance-based visualizations of confidence scores with interactive example-based explanations (Figure \ref{fig:interface_during}). 
Using the dataset of 15 post-stroke survivors and 10 healthy participants, we developed an AI-based decision-support system that employs a feed-forward neural network to assess the quality of motion. Grounded in prior findings that health professionals value interpretable, important feature explanations \cite{lee2021human}, our system also presents important feature explanations to assist users' rehabilitation assessment tasks (Figure \ref{fig:interface_during}). To support more trustworthy and calibrated decision-making, we implemented interactive components designed to enhance users' understanding and use of confidence scores (Figure \ref{fig:interface_before} and \ref{fig:interface_during}). Specifically, our interactive system allows a user to explore and set a threshold of confidence scores to delegate cases to AI, review AI performance on delegated cases using a held-out dataset, and determine which cases will be delegated to AI (Figure \ref{fig:interface_before}). 

Instead of just presenting a numerical confidence score, our interactive system identifies the $k$-nearest neighborhoods of a given case and visualizes their embedding spaces along with the centroid of class labels. Users can interact with this visualization by hovering over individual data points to access contextual information about each neighbor, such as whether the neighbor is a post-stroke survivor, the AI model's performance, and the level of agreement among therapists on the neighbor post-stroke survivor. We hypothesize that our interactive exploration and distance-based visualization of confidence scores with example-based explanations can support more analytical user engagement with AI outputs and confidence scores \cite{zhang2020effect,prabhudesai2023understanding,lee2023understanding}, fostering better-calibrated reliance on AI for human-AI collaborative decision-making.

Among various uncertainty quantification techniques \cite{smith2013uncertainty,hendrycks2016baseline,corbiere2019addressing,gal2016theoretically,kendall2017uncertainties,jiang2018trust,papernot2018deep,van2020uncertainty}, we empirically found that a distance-based approach \cite{kendall2017uncertainties,jiang2018trust,papernot2018deep,van2020uncertainty} was more effective to identify uncertain cases (Section \ref{section:uncertainty_quantification}). Specifically, our distance-based approach utilizes the vector spaces on the processed layers of the AI model and quantifies uncertainty by computing the distances between a given case and the centroids of each class in the embedding space.

To evaluate the effectiveness of our interactive exploration and distance-based visualization of confidence scores with example-based explanations, we conducted a 2x2 mix-design experiment with 9 domain experts (i.e. therapists with experience in stroke rehabilitation) and 20 novices (i.e. other health professionals without/with limited experience in stroke rehabilitation and students in medicine, nursing, and therapy) for AI-assisted decision-making. 

Our results showed that our interactive exploration of confidence score thresholds for task delegation to AI (Figure \ref{fig:interface_expconf}) reduced participants' overreliance on wrong AI outputs by achieving a 3.58\% lower overreliance rate with numerical confidence scores and a 1.85\% lower overreliance rate with distance-based confidence scores and interactive example explanations, compared to those who did not explore confidence score thresholds. In addition, our distance-based confidence scores with interactive example-based explanations significantly improved decision-making, leading to an 8.20\% higher rate of correct decisions, a 7.15\% higher rate of changing their decisions to correct ones, and a 7.14\% lower rate of incorrect changes after reviewing AI outputs, compared to those reviewing numerical confidence scores ($p<0.01$).

Taken together, our work contributes to the growing body of research on human-AI collaborative decision-making by examining how to make uncertainty quantification more accessible and actionable for users without AI expertise. We demonstrate the potential of interactive explorations and functionalities, such as reviewing AI/ML components (e.g. confidence scores) to encourage more analytical engagement with AI/ML components for health professionals and decision tasks. In addition, we highlight the challenges in designing educational and interpretive mechanisms that foster critical review of AI outputs while minimizing cognitive and interactional burdens for effective human-AI collaborative decision-making.

\section{Related Work}
\subsection{Human-AI Collaborative Decision Making}
Recent research on AI has demonstrated its potential to replicate expert-level decision-making 
\cite{lee2019learning,nam2019development,singh2018deep,esteva2017dermatologist}.\\ 
AI-based decision-support systems are increasingly being explored to reduce decision-makers' cognitive burden and improve their decision outcomes \cite{lai2021towards} in various domains, such as healthcare \cite{cai2019human,lee2021human,kim2024much}, law \cite{lima2021human}, and other public services \cite{de2020case,stapleton2022imagining,kawakami2022improving,kuo2023understanding}. However, fully automated approaches to AI-assisted decision-making remain problematic in high-stakes contexts due to ethical, legal, and accountability concerns. Thus, there is growing interest in collaborative approaches between humans and AI \cite{lai2021towards,lee2021human,salimzadeh2024dealing,cai2019hello,cai2019human,holstein2023toward}.

Human-AI collaborative approaches aim to leverage the strengths of humans and AI to achieve the complementary and enhanced decision outcomes \cite{lai2021towards,lee2021human,salimzadeh2024dealing,cai2019hello,cai2019human,holstein2023toward,buccinca2021trust}. Prior research has explored ways to support human decision-making by providing data-driven insights (e.g. retrieving similar cases from previously diagnosed patients \cite{cai2019human,chen2023understanding,lee2024interactive}, identifying important input features \cite{lee2021human,wang2021explanations}, and generating counterfactual explanations \cite{lee2023understanding,zhang2022towards}). These techniques have shown potential to enhance domain experts' accuracy and efficiency of decision making into practice \cite{cai2019human,lee2021human,beede2020human,wang2021brilliant,lee2023understanding}. 

Despite these advancements, the adoption of AI-based decision-support systems remains limited due to the issue of user acceptance and trust \cite{sutton2020overview,khairat2018reasons,kohli2018cad,wang2021brilliant}. Domain experts (e.g. clinicians, health professionals) often lack technical backgrounds and may struggle to understand the system's intended use, functionalities, or capability \cite{maddox2019questions,sutton2020overview,lee2024improving}. This lack of clarity can lead to skepticism, resistance, and eventual abandonment of such tools \cite{khairat2018reasons}. In addition, the opaque nature of AI systems with complex algorithms further exacerbates these concerns \cite{sutton2020overview,rajpurkar2022ai,cai2019human}. When AI operates as a black box, users may find it difficult to understand how specific AI outputs are generated \cite{london2019artificial,rajpurkar2022ai,cai2019human} and decide whether or not they should rely on AI outputs or not \cite{bussone2015role,buccinca2021trust,lee2023understanding,ma2023should,salimzadeh2024dealing}. 

In this work, we focus on human-AI collaborative decision-making in the context of physical stroke rehabilitation assessment. Specifically, we investigate how to enhance user's understanding of AI uncertainty quantification and examine the impact of two features: (i) interactive exploration of confidence score thresholds and (ii) distance-based confidence scores with interactive example-based explanations, exploring how these features influence non-technical users' reliance on AI for human-AI collaborative decision-making. 

\subsection{Explainable AI and Uncertainty Quantification for Human-AI Collaborative Decision-Making}
To make AI more trustworthy \cite{floridi2019establishing,liang2022advances,araujo2020ai,jacovi2021formalizing} and deployable in practice, researchers have proposed various design considerations and factors of trustworthy AI \cite{wing2021trustworthy,vashney2022trustworthy,floridi2019establishing,toreini2020relationship}. Even if there is still limited consensus on a unified definition of trustworthy AI \cite{gille2020we}, multiple frameworks have identified key components of trustworthy AI. In this work, we focus on the understandability aspects of AI that describe whether users can comprehend AI model's outputs and behavior \cite{vashney2022trustworthy,toreini2020relationship}. 

To enhance user's understanding of AI models, researchers have explored various computational techniques, such as explainable AI \cite{arya2019one,wang2019designing,lakkaraju2020explaining,abdul2018trends,preece2018asking} and uncertainty quantification \cite{hendrycks2016baseline,corbiere2019addressing,gal2016theoretically,kendall2017uncertainties,jiang2018trust,papernot2018deep,van2020uncertainty,zhang2024evaluating}. These computational approaches aim to either explain why an AI model generates a certain output \cite{arya2019one,wang2019designing,lakkaraju2020explaining,abdul2018trends,preece2018asking} or to quantify the confidence of an AI output, assisting humans to determine when to trust and rely on AI outputs. In the following subsections, we summarize the related work on explainable AI and uncertainty quantification techniques, highlighting how our approach builds upon and differentiates from prior research work. 

\subsubsection{Explainable AI}
Researchers have explored explainable AI (XAI) techniques to address the challenge of understanding why an AI model generates specific outputs \cite{preece2018asking,arya2019one,wang2019designing,lakkaraju2020explaining}. XAI techniques can be broadly categorized into either making inherently interpretable models \cite{lakkaraju2020explaining,rudin2019stop} or posthoc XAI techniques. Inherently interpretable models refer to AI/ML models whose internal mechanisms are directly understandable, such as rule-based models or linear regressions \cite{lakkaraju2020explaining,rudin2019stop}. Post-hoc XAI techniques analyze and provide explanations for the overall or instance-specific behaviors of AI \cite{lakkaraju2020explaining}. Among various posthoc XAI techniques, this work focuses on two widely used local XAI techniques: feature importance and prototype/example-based explanations. Feature importance explanations describe how individual input features contribute to a model's output \cite{gilpin2018explaining,ribeiro2016should,mundhenk2019efficient,lakkaraju2020explaining}. Prototype/example-based explanations aim to identify samples that are the most relevant or influential to an AI output \cite{gilpin2018explaining,cai2019effects,chen2023understanding,lee2024interactive}.

\subsubsection{Uncertainty Quantification}
Uncertainty quantification refers to the process of estimating the uncertainty or confidence associated with the outputs of an AI model \cite{smith2013uncertainty}. In this work, we investigate uncertainty quantification techniques to assist users in better assessing when AI outputs can be trusted or not for human-AI collaborative decision-making \cite{zhang2024evaluating,hullman2018pursuit}. 
Most commonly used methods of uncertainty quantification include the following approaches \cite{caldeira2020deeply}: (1) the maximum class probability \cite{hendrycks2016baseline}, (2) a model- and task-agnostic approach of confidence estimation \cite{corbiere2019addressing}, (3) a Bayesian approach \cite{gal2016theoretically,kendall2017uncertainties}, and (4) distance-based approaches \cite{jiang2018trust,papernot2018deep,van2020uncertainty}.

One line of work in uncertainty quantification focuses on transforming classifier outputs into confidence scores  \cite{hendrycks2016baseline,corbiere2019addressing}. A commonly used method is the maximum class probability \cite{hendrycks2016baseline}, which estimates confidence by taking the maximum probability score from the model's output. In contrast, Corbière et al. \cite{corbiere2019addressing} proposed a new target of model confidence called, a true class probability. As the true class is unknown during inference, Corbière et al. investigated to learn a model- and task-agnostic approach to learn and approximate the true class probability \cite{corbiere2019addressing}. 

Another prominent direction involves Bayesian approaches to uncertainty estimation \cite{gal2016theoretically,kendall2017uncertainties} that has gained a lot of attention due to the connection between an efficient stochastic technique (e.g. dropout \cite{gal2016theoretically}) and variational inference in Bayesian neural networks. Specifically, Gal and Ghahramani showed that applying Monte Carlo Dropout, which produces multiple stochastic forward predictions/passes of a network to estimate a posterior distribution over model outputs \cite{gal2016theoretically}.

In addition, a distance-based approach of uncertainty quantification utilizes the vector space on processed layers of an AI model and computes the distances with the class labels for uncertainty estimations \cite{jiang2018trust,papernot2018deep,van2020uncertainty}. Researchers have experimentally shown that a distance-based approach using the intermediate representations of an AI model is more robust to outliers and leads to better-calibrated uncertainty estimations \cite{jiang2018trust,papernot2018deep}.

In this work, we empirically investigate which uncertainty quantification techniques perform well to identify uncertain cases that require human reviews. Beyond comparing performance, we also examine the effect of distance-based visualization of confidence scores along with interactive example-based explanations, which present the distances of embedding spaces of input data and class centroids to assist user's understanding of uncertainty scores and making more informed AI-assisted decision-making. 

\subsection{Studies on Human-AI Collaborative Decision-Making and Overreliance on AI}
With the growing interest in XAI and uncertainty quantification, researchers have investigated the effect of presenting information about AI outputs using XAI and uncertainty quantification for various decision-making tasks (e.g. AI-advised image labeling \cite{zhang2024evaluating}, trip-planning \cite{salimzadeh2024dealing}, student admission \cite{cheng2019explaining}, deception detection \cite{lai2019human}, cancer diagnosis \cite{cai2019human}, and stroke rehabilitation assessment \cite{lee2021human}). 

Empirical studies have surfaced varying perspectives on the effectiveness of providing additional information to humans and have revealed challenges in fostering appropriate reliance on AI. While some studies demonstrated that presenting AI explanations does not affect users' trust \cite{cheng2019explaining}, other studies found that AI explanations could increase users' trust in AI and lead to overreliance on `wrong' AI outputs \cite{bussone2015role,bansal2021does,buccinca2021trust,lee2023understanding}. 
In addition, Prabhudesai et al. described the potential of presenting uncertainty about AI outputs to support users' analytical thinking and reduce over-reliance on AI \cite{prabhudesai2023understanding}. However, recent work has also shown that presenting uncertainty quantification alone does not necessarily guarantee improved decision-making \cite{zhang2024evaluating}. These failures of presenting additional information on AI outputs including AI explanations and uncertainty quantification may stem from the way information is communicated, particularly to non-technical users, who may struggle to interpret or act on it effectively \cite{suresh2021beyond}. Along these lines, researchers highlight the importance of effectively communicating uncertainty to increase critical evaluation of AI outputs \cite{zhang2020effect,prabhudesai2023understanding} and foster more trustworthy human-AI interaction \cite{amershi2019guidelines,lai2021towards,lee2024improving}.

Even if prior work has investigated the usefulness of confidence scores during AI-assisted decision-making phase \cite{zhang2020effect,prabhudesai2023understanding,zhang2024evaluating}, few studies have explored the utility of uncertainty before the AI-assisted decision-making phase to identify cases to delegate to AI and how to effectively present uncertainty quantification to increase analytical reviews and engagement with AI outputs \cite{zhang2020effect,prabhudesai2023understanding}. 

Building upon the prior literature, we investigate to understand the utility of uncertainty both before and during AI-assisted decision-making phase, focusing on its impact on users' reliance on AI for human-AI collaborative decision-making. To this end, we developed an interactive AI-based decision-support system designed to enhance user's understanding and utility of uncertainty scores with the following two key features: (1) users can explore a threshold of confidence scores to delegate cases to AI and learn how well an AI model performs on the delegated cases using a held-out dataset (Figure \ref{fig:interface_before}); and (2) users can review distance-based visualizations of confidence scores using embedding data and interactive example-based explanations instead of relying solely on a numerical presentation of confidence scores (Figure \ref{fig:interface_during}). We then conducted a 2x2 mix-design experiment with health professionals and students in medicine/health in the context of physical stroke rehabilitation assessment. Our analysis examines the effectiveness of these interactive features in shaping users' reliance on AI.

\section{Study Design}
This work aims to investigate the utility of uncertainty scores for task delegation and human-AI collaborative decision-making (i.e. physical stroke rehabilitation assessment). Building upon prior research highlighting the importance of effectively communicating uncertainty \cite{zhang2020effect,prabhudesai2023understanding,salimzadeh2024dealing} and promoting critical evaluation of AI outputs \cite{buccinca2021trust,lee2023understanding,prabhudesai2023understanding}, we explored how domain-specific users can leverage uncertainty in AI outputs to identify cases suitable for delegation to AI and reduce overreliance on incorrect AI outputs \cite{kaur2020interpreting,wang2021explanations,lai2021towards}. Instead of presenting uncertainty scores numerically, we hypothesize that our distance-based visualization of uncertainty scores can improve users' understanding of AI uncertainty and increase more critical engagement with AI outputs, ultimately supporting more effective human-AI collaborative decision-making. 
 %to reduce overreliance on AI for 

To explore our research questions, we first evaluated the performance of various uncertainty quantification (UQ) techniques for identifying uncertain cases to delegate to AI. We then conducted a user study with health professionals and students in medicine or health to examine the utility of uncertainty scores to determine cases to delegate to AI (Figure \ref{fig:interface_before}) and decide the reliance on AI outputs during decision-making (Figure \ref{fig:interface_during}). For the presentation of uncertainty scores, we analyzed the effect of visualizing embedding spaces of data points and class centroids, comparing it to a baseline system with a numerical presentation of uncertainty (Figure \ref{fig:interface_during}).

\subsection{Study Context of Physical Stroke Rehabilitation Assessment}
In this work, we focus on decision-making tasks of assessing the quality of motion of post-stroke survivors. Building upon prior research in human-AI decision-making on physical stroke rehabilitation assessment \cite{lee2021human}, we specified an upper limb exercise and the performance components of physical stroke rehabilitation assessment. Specifically, we utilize an exercise, in which post-stroke survivors have to raise their wrist to their mouth, simulating the motion of drinking water. The performance components of rehabilitation assessment are composed of `Range of Motion (ROM)' that evaluates how well a post-stroke survivor reaches the target position of an exercise and `Compensation' that assesses if a post-stroke survivor involves any unnecessary, compensatory joint movements to perform an exercise (e.g. raising shoulder or leaning backward).

\subsection{System Implementations}
Our AI-based system for physical stroke rehabilitation assessment (Figure \ref{fig:interface_before} and \ref{fig:interface_during}) utilizes a feed-forward neural network (NN) model to classify post-stroke survivor's quality of motion and presents AI predictions on rehabilitation assessment along with confidence scores \cite{amershi2019guidelines} and interactive AI explanations. Prior to AI-assisted decision-making, the system allows users to explore different thresholds of confidence scores and review AI performance on delegated cases from a held-out dataset (Figure \ref{fig:interface_expconf}). For instance, if a user specifies a threshold of 60\%, the system informs that the AI model has 54.55\% accuracy on the cases exceeding this threshold (Figure \ref{fig:interface_expconf}). Once the user specifies a threshold of confidence scores to delegate cases to AI (Figure \ref{fig:interface_expconf}), the user can review and confirm which cases are appropriate for delegation to the AI and which require further human review (Figure \ref{fig:interface_taskdelegate}).

The system also provides 1) interactive example-based explanations to support users' understanding of AI confidence scores by visualizing the embedding representations of an input case, class centroids, and information about its nearest neighbors and 2) important feature explanations for assisting user's decision-making on rehabilitation assessment (Figure \ref{fig:interface_during}). Our system interface was built with Python, Flask \cite{grinberg2018flask}, and HTML/Javascript libraries. 

\subsubsection{Dataset}
We utilize the \textit{``Bring a cup to the mouth''} upper-limb exercise dataset from 15 post-stroke survivors with diverse functional abilities \cite{lee2019learning} and 10 healthy participants, who performed correct motions and acted out incorrect motions of post-stroke survivors \cite{hun2023design}. The dataset contains (1) 300 videos of 15 post-stroke survivors, performing ten trials of the exercise using their unaffected and affected arms by stroke and 60 videos of 11 healthy participants, performing one normal/correct movement and five acted-out incorrect movements, (2) estimated joint positions of their exercise motions using a Kinect sensor v2, and (3) the annotations by the expert therapist, who utilized a clinically validated assessment tool \cite{gladstone2002fugl} for post-stroke survivors during the recruitment and another therapist without interactions with them. 
%For the annotations, therapists individually watched the videos of post-stroke survivors without reviewing any AI outputs.

\subsubsection{AI Model}\label{sect:ai_model}
Building on previous research in rehabilitation assessment \cite{lee2019learning}, we processed the estimated joint positions of post-stroke survivors' exercises to extract various kinematic features (Appendix \ref{appendix:feature_extract}). In addition, following the previous findings that show the competitive performance of a feed-forward NN model for assessing the quality of motion \cite{lee2019learning}, we implemented it using the Pytorch libraries \cite{paszke2019pytorch} (Appendix \ref{appendix:results_ml}). 
%For the labels, we utilized the labels by the expert therapist, who conducted the clinically validated assessment test with post-stroke survivors as ground truth. 
%For a feed-forward NN model with the parameters $\textbf{W}$, 
The final model architectures and learning rates of our models are three layers of 256 hidden units and a learning rate of 0.00500 for the `ROM' performance component and three layers of 32 hidden units, the last layer with 64 hidden units, and a learning rate of 0.0005 for the `Compensation' performance component. Using the leave-one-subject-out cross-validation, our AI model achieved an average of 81.97\% F1-score to replicate therapist's assessments using data from post-stroke survivors and healthy participants and an average of 82.76\% F1-score using post-stroke survivors' data (Appendix. Table \ref{tab:results_mlmodels}).

\subsubsection{Uncertainty Quantification}\label{section:uncertainty_quantification}
In this work, we explore several techniques to estimate the confidence of an AI model prediction: the maximum class probability \cite{hendrycks2016baseline}, a model- and task-agnostic approach of confidence estimation \cite{corbiere2019addressing}, a Bayesian approach \cite{gal2016theoretically,kendall2017uncertainties}, and distance-based approaches \cite{jiang2018trust,papernot2018deep,van2020uncertainty}. In general, a higher confidence score indicates that the AI model is more certain about its prediction, while a lower score suggests greater uncertainty. In other words, we assume that successful, right AI predictions tend to be associated with high confidence scores, whereas erroneous, wrong AI predictions are likely to yield low confidence scores. 

We briefly described the implementation details in Appendix \ref{appendix:uq}. One of the most commonly used approaches of a confidence criterion is \textbf{maximum class probability} \cite{hendrycks2016baseline}, which computes the softmax probability of the predicted class $\hat{Y}$.
%$MCP(\textbf{X}) = \max_{k \in \mathcal{Y}} P(Y = k |\textbf{W, X})$ 
\begin{comment}
\begin{equation}
MCP(\textbf{X}) = \max_{k \in \mathcal{Y}} P(Y = k |\textbf{W, X})
\end{equation}
\end{comment}
%= \hat{c}(\textbf{X}, \theta)$ 

For a \textbf{confidence estimation approach}, we followed the approach, which proposes to utilize True Class Probability (TCP) associated with the true class $Y*$ and learn a model to estimate this TCP in the context of predicting a model failure \cite{corbiere2019addressing}. 

For \textbf{Bayesian approaches}, we adopted Monte Carlo Dropout (MCDropout) as proposed by Gal and Ghahramani \cite{gal2016theoretically}, which estimates the posterior distribution of model predictions by performing multiple stochastic forward passes through the network during inference. 

For \textbf{distance-based approaches}, we explored NN-based and radial basis function (RBF)-based distance approaches \cite{jiang2018trust,papernot2018deep,van2020uncertainty}. Specifically, the {NN-based} distance approach leverages the intermediate representations of a feed-forward NN model to compute the centroid of classes and input data \cite{jiang2018trust,papernot2018deep} or an RBF network with a set of feature vectors corresponding to the different classes (centroids) \cite{van2020uncertainty}. After computing the centroids of classes and input data using either an NN-based or an RBF-based model, we measured the distance between the model output and the nearest centroid to estimate uncertainty \cite{jiang2018trust,papernot2018deep,van2020uncertainty}. Specifically, given $K$ classes, we compute the distance $d$ between the input and the centroids of each class (among $K$ classes), normalize it by the maximum distance, $\hat{d} = (1 - \frac{d}{d_{max}})$, and apply a softmax function to obtain normalized confidence scores: $\frac{e^{\hat{d}}}{\sum_{j=0}^{K}e^{\hat{d_j}}}$.

Among the UQ approaches, maximum class probability, the Bayesian approach using MCDropout, and the RBF-based distance approach do not involve calibration of confidence scores. In contrast, the confidence estimation approach learns to predict calibrated confidence scores by training a separate model. Our feed-forward NN-based distance approach incorporates calibration by normalizing the distance measure with the maximum observed distance and applying a softmax function to produce confidence scores. 

\textbf{Evaluations of Uncertainty Quantification Approaches}:\\
In our work, we assume that a trained AI/ML model generates confidence scores for its predictions but lacks an appropriate threshold to identify uncertain cases. To evaluate various uncertainty quantification (UQ) approaches, we explored thresholds ranging from 0 to 1 with an increment of 0.05. For cases with confidence scores below a given threshold, we replaced the model's predicted labels of uncertain cases with the therapist's labels, allowing us to evaluate the quality of UQ approaches for identifying uncertain cases. An effective UQ approach should accurately identify uncertain cases, replace their predictions with therapist's assessments, and ultimately achieve higher performance on decision-making tasks.
%When an uncertainty approach can perform well in identifying uncertain cases, the approach will replace those with therapist's assessments and have higher performance on decision-making tasks and vice versa. 

We summarize the average performances of various uncertainty quantification approaches (i.e. maximum class probability, feed-forward NN or RBF-based distance approaches, a confidence network, and a Bayesian approach) in identifying uncertain cases to replace with expert annotations across various thresholds (Table \ref{tab:results_uqmodels_simple}). 

%\input{tables/results_uq}

% \usepackage{graphicx}
\begin{table}[]
\centering
\caption{Results of Identifying Uncertain Cases using Machine Learning Models and Uncertainty Techniques. The Feed-Forward Neural Network models using a distance-based uncertainty quantification approach achieved the highest performance.}
\label{tab:results_uqmodels_simple}
\resizebox{\columnwidth}{!}{%
\begin{tabular}{ccccc} \toprule
\multicolumn{2}{c}{\textbf{UQ Models}} &
  \textbf{\begin{tabular}[c]{@{}c@{}}All\\ (Post-Stroke + Healthy)\end{tabular}} &
  \textbf{Post-Stroke} &
  \textbf{Healthy} \\ \midrule
\begin{tabular}[c]{@{}c@{}}Feed-Forward\\ NN \cite{hendrycks2016baseline}\end{tabular} &
  \begin{tabular}[c]{@{}c@{}}Maximum Class\\ Probability (MCP)\end{tabular} &
  82.86 &
  85.33 &
  79.14 \\ \midrule
\begin{tabular}[c]{@{}c@{}}Confidence\\ Network \cite{corbiere2019addressing}\end{tabular} &
  \begin{tabular}[c]{@{}c@{}}Estimating\\ True Class\\ Probability\end{tabular} &
  79.94 &
  83.13 &
  75.17 \\ \midrule
\begin{tabular}[c]{@{}c@{}}MCDropOut\\ Network \cite{gal2016theoretically}\end{tabular} & Bayesian       & 72.92 & 78.80 & 64.10 \\ \midrule
RBF Network \cite{van2020uncertainty}                                                 & Distance-based & 76.93 & 86.17 & 60.37 \\ \midrule
\begin{tabular}[c]{@{}c@{}}Feed-Forward\\ NN\end{tabular} &
  Distance-based &
  \textbf{93.63} &
  \textbf{91.77} &
  \textbf{91.53} \\ \bottomrule
\end{tabular}%
}
\end{table}

Overall, our \textbf{feed-forward NN-based distance} approach outperformed all other approaches, achieving the highest average F1-score of 93.63\% in assessing the quality of motion after replacing uncertain AI predictions with expert's labels. Notably, UQ approaches generally performed poorly on data from healthy participants due to the presence of diverse, acted-out incorrect motions that differ from those of post-stroke survivors. Despite this challenge, our feed-forward NN-based distance approach demonstrated robust performance on both in-distribution and out-of-distribution data. In addition, this NN-based distance approach provides interpretable insights by enabling visualization of embedding spaces of data and class centroids, assisting to explain the resulting confidence scores. 

%The performance of other UQ approaches can be found in Appendix. Table \ref{tab:results_uqmodels}. 

\subsubsection{Interactive Example-based Explanations for Confidence Scores}\label{section:example_explanations}
To enhance user's understanding of confidence scores and analytical reviews on AI outputs, we propose interactive example-based explanations (Figure \ref{fig:interface_during}). Our interactive example-based explanations \cite{lee2024interactive} visualize the embedding space \cite{boggust2022embedding} of a given case (a black point in Figure \ref{fig:interface_during}) along with its $k$-nearest neighbors and class centroids (purple, orange, and green points in Figure \ref{fig:interface_during}). Users can explore the number of $k$-nearest neighbors and click individual embedding data to review the images of a neighbor post-stroke survivor (`Relevant Images' in Figure \ref{fig:interface_during}). Also, users can hover over data to quickly review the benchmarkable information of a neighbor via tooltips: the status of a neighbor post-stroke survivor, the AI model's performance, and therapist's agreement level on the neighbor post-stroke survivor (RelevantInfo in Figure \ref{fig:interface_during}).

%%%%%%%
%Unlike the traditional numerical presentation of a confidence score (e.g. ``Confidence: 91\%'') \cite{amershi2019guidelines,lee2021human},  We hypothesize that these interactive explanations can enhance users' understanding of confidence scores and decisions about relying on AI for effective human-AI collaborative decision-making.

We experimented with the representations of input and intermediate layers of the feed-forward NN models, applying various dimensionality reduction techniques, such as Principal Component Analysis (PCA) \cite{abdi2010principal}, Uniform Manifold Approximation and Projection (UMAP) \cite{becht2019dimensionality}, and t-distributed Stochastic Neighbor Embedding (t-SNE) \cite{van2008visualizing}. Given embedding data from these techniques, we implemented a $K$-nearest neighbor classifier with various $k$ ranging from 5 to 30, using either cosine or Euclidean distance metrics \cite{turney2010frequency}.  Based on the experimental results (Appendix. Table \ref{tab:results-drt}), we utilized embedding data using the t-SNE from the first input layers of the feed-forward NN models and applied Euclidean distance metric.

\subsubsection{Feature-based Explanations for Decision-Support}\label{section:feature_explanations}
Building on prior research that highlights therapists' preferences for reviewing feature-based explanations on rehabilitation assessment tasks \cite{lee2020co}, we implemented a feature-based explanation to assist users' decision-making. For feature-based explanations (Figure \ref{fig:interface_during}), we identified user-specific important features using the feed-forward NN models and the SHAP library \cite{NIPS2017Shap}. To prevent cognitive overload \cite{amershi2019guidelines}, we displayed only the top three features. In alignment with therapists' practices to compare the unaffected and affected side of post-stroke survivors, we visualized these features using a radar chart to facilitate intuitive comparison and assist users' decision-making on rehabilitation assessment \cite{lee2020co}.

\subsection{Experimental Designs}\label{sect:experiment_designs}
We describe our experimental designs including experimental groups and conditions, participants, protocol, and evaluation metrics in detail. All study materials and procedures were reviewed and approved by the Institutional Review Board (IRB).

\subsubsection{Experimental Groups \& Conditions}
We defined two participant groups and two experimental conditions to explore the effectiveness of our interactive exploration and example-based explanations on confidence scores for user's decision-making of rehabilitation assessment tasks. Two participant groups were designed to understand the effect of interactive explorations of confidence scores for task delegation to AI. The two conditions were to evaluate how our interactive visualization of confidence scores with example-based explanations influences users' reliance on AI during human-AI collaborative decision-making.

\begin{itemize}
    \item \textbf{``NoExp''}: Baseline, controlled group of participants, who did not explore to set a threshold of confidence scores and directly determine which cases to delegate to AI (Figure \ref{fig:interface_taskdelegate}).
    \item \textbf{``Exp''}: Experimental group of participants, who explored to set a threshold of confidence scores (Figure \ref{fig:interface_expconf}) before determining which cases to delegate to AI (Figure \ref{fig:interface_taskdelegate}).
\end{itemize}

\begin{itemize}
    \item \textbf{``Numerical Confidence Scores''}: Baseline, controlled condition of the AI system that presents videos of post-stroke survivors' exercises, AI predicted assessment scores, and important feature explanations. Confidence scores are presented \textbf{numerically, without interactive example-based explanations} (Figure \ref{fig:interface_during}).
    \item \textbf{``Distance-based Confidence Scores''}: Experimental condition of the AI system that includes all functionalities of the baseline condition along with the \textbf{distance-based visualization of confidence scores with interactive example-based explanations} (Figure \ref{fig:interface_during}).
\end{itemize}

Building on prior research that described therapists' preferences to review feature-based explanations \cite{lee2020co}, we included important feature explanations by default to assist users in completing rehabilitation assessment tasks \cite{wang2019designing}. During the study, we referred to two systems as ``Condition A'' and ``Condition B'' to minimize potential bias. In this paper, we referred to these conditions as ``Numerical Confidence Scores" and ``Distance-based Confidence Scores''.

\subsubsection{Participants}
We recruited 29 participants for our study through an advertisement sent to hospital staff, mailing lists, and the research team's professional contacts. Detailed demographics of the participants are provided in Appendix. Table \ref{tab:participants_details}. Of 29 participants, we have 9 \textbf{domain experts}, therapists with over one year of experience in stroke rehabilitation: 6 occupational therapists and 3 physiotherapists. The remaining 20 participants were categorized as \textbf{novices}: 10 health professionals (e.g. 3 health/therapist assistants, 4 healthcare workers, 2 nurses, and 1 doctor) with limited experience of stroke rehabilitation and 10 students majoring in medicine (3), nursing/health (4), or occupational therapy (3). 
%Participated domain experts, therapists, work in various settings: 3 therapists from acute care, 3 therapists from inpatient rehabilitation, 3 therapists from outpatient clinics or skilled nursing facilities (Appendix. Table \ref{tab:participants_details}). Among 9 therapists, we have 6 occupational therapists, who support post-stroke survivors to better engage in their daily activities and 3 physiotherapists, who support to maintain and improve post-stroke survivors' physical impairments from biomechanical perspectives. 
For data analysis, we excluded two novice participants due to their extremely short interaction time with the interface or incomplete log data. Overall, we had 15 participants of `NoExp' (6 therapists and 9 novices) and 12 participants of `Exp' (3 therapists and 9 novices).

\subsubsection{Protocol}
We conducted a 2x2 mixed-design experiment to understand the effects of interactive exploration for task delegation to AI and distance-based visualizations of confidence scores with interactive example-based explanations for users' reliance on AI during decision-making. After providing informed consent, participants were randomly assigned to either no exploration of a threshold (NoExp) or the exploration of confidence score thresholds for task delegation (Exp). In addition, all participants experienced two within-subject conditions in a counterbalanced order: using AI with important feature explanations and numerical confidence scores (Condition A), and using AI with important feature explanations and distance-based confidence scores with interactive example-based explanations (Condition B) (Figure \ref{fig:interface_taskdelegate}).

After reviewing a tutorial on the AI system and study procedures, each participant completed either four (NoExp) or five (Exp) sub-tasks. First, all participants independently assessed 14 post-stroke survivors' videos without reviewing AI outputs and explanations. Following their initial assessment, participants in the NoExp group directly reviewed the confidence scores of AI outputs and selected which cases to delegate to AI (Figure \ref{fig:interface_taskdelegate}). In contrast, participants in the Exp group explored a threshold for confidence scores (Figure \ref{fig:interface_expconf}) while reviewing AI performance on delegated cases from a held-out dataset before deciding which cases to delegate to AI (Figure \ref{fig:interface_taskdelegate}). After confirming the cases to delegate to AI, participants reviewed AI outputs and explanations with numerical confidence scores (Condition A) and distance-based confidence scores with interactive example-based explanations (Condition B) to complete the assessment of the assigned cases.
%After completing the assessment tasks using each condition (Condition A or B), the participants completed the usability questionnaires on each condition. 
Finally, after completing all assessment tasks, participants completed the overall, post-study questionnaires. All participants received fixed compensation based on the rate recommended by the domain experts. 

Each participant completed a subtask consisting of 14 assessment cases per condition/system. To examine the impact of interactive explorations and visualization of confidence scores with example-based explanations on users' overreliance on `wrong' AI outputs, we used our trained feed-forward NN model (Section \ref{sect:ai_model}) to select the cases of post-stroke survivors. Specifically, each condition included 10 cases with correct AI outputs and 4 with incorrect, reflecting approximately 71\% accuracy - similar to the AI model's performance. 
 We counterbalanced the assignment of cases across conditions and randomized both the order of the two conditions and the sequence of presenting assigned cases. For the NoExp group, we conducted a grid-search over confidence thresholds ranging from 0 to 1 to empirically determine a default threshold (i.e. 40\%) to identify high confidence scores. Participants then reviewed these scores to identify cases to delegate to AI. For Numerical Confidence Scores, we utilized the maximum class probability from our feed-forward NN model. For the Distance-based Confidence Scores, we applied an NN-based distance approach to compute confidence scores. 

\subsubsection{Data Analysis Metrics}
Building upon previous research on human-AI decision-making \cite{lai2021towards,cai2019human,lee2023understanding}, we employed several data analysis metrics: 1) the ratio of Right decisions using  ground truth labels, 2) the ratio of Changed decision after reviewing AI outputs and explanations, 3) the ratio of Right decisions before/after reviewing AI outputs and explanations, 4) the ratio of decisions that changed to Right (ChangedRight), 5) the duration of decision-making tasks, and 6) post-study questionnaires on which system components help participants to validate the system's competence and determine their reliance on it (Appendix. \ref{appendix:study_metrics}).

\section{User Study Results}\label{sect:results_study}
In this section, we describe the user study results on our data analysis metrics. We refer to the decision-making of participants who did not review AI outputs and explanations as \textbf{{``Human''}} and decision-making of those who reviewed AI outputs and explanations as \textbf{{``Human + AI''}}. We categorize all participants (\textbf{{``All''}}) into four groups: 
(i) \textbf{{NoExp}} participants, who did not explore a threshold of confidence scores before delegating cases to AI, (ii) \textbf{{Exp}} participants, who explored a threshold of confidence scores for task delegation to AI, (iii) \textbf{{``TPs''}} domain experts, therapists, with over one year of experience in stroke rehabilitation, and (iv) \textbf{{``NVs''}} novices, who are health professionals or students majoring in medicine/health/occupational therapy and have less than one year of experience with stroke rehabilitation.

%In addition, we refer to Condition A, in which participants interact with our AI system with important feature explanations and numerical confidence scores \textbf{\textit{``Numerical Confidence Scores''}} and Condition B, where participants interact with our AI system with important feature explanations and distance-based confidence scores along with interactive example-based explanations \textbf{\textit{``Distance-based Confidence Scores''}}.

To evaluate participant performance, we compared differences between {{Human}} and {{Human + AI}} over two conditions (i.e. {{Numerical Confidence Scores}} and {{Distance-based Confidence Scores}}) and different groups of participants ({{NoExp}} vs. {{Exp}} and {{TPs}} vs. {{NVs}}). For other data analysis metrics, we analyzed the differences over two conditions and four different groups of participants. For each metric, we computed the descriptive statistics and conducted a significance test, in which we first checked the normality of data on each metric using the Kolmogorov–Smirnov test \cite{massey1951kolmogorov}. If the data followed a normal distribution, we utilized paired t-tests. Otherwise, we applied the Wilcox significance tests \cite{cuzick1985wilcoxon} to analyze data from different conditions and participant groups respectively.

\subsection{Ratio of `Right' decisions}
Overall, all participant groups using Distance-based Confidence Scores with interactive example-based explanations achieved higher ratios of Right decisions than those using only Numerical Confidence Scores (Table \ref{tab:results_ratio_right}). Specifically, the Right decision ratio with \textbf{Distance-based Confidence Scores with interactive example-based explanations was 74.34\%, which is 8.2\% increase over 66.14\% observed with Numerical Confidence Scores} ($p<0.01$).

% \usepackage{graphicx}
\begin{table}[htp]
%\caption{Ratio of `Right' decisions after reviewing AI outputs (`Human + AI'): all participants (All), participants without exploring a threshold of confidence scores before task delegations (NoExp), participants with exploring a threshold of confidence scores for task delegations (Exp), domain experts, therapists (TPs), and other health professionals and students (NVs)}
\caption{Ratio of `Right' decisions after reviewing AI outputs (Human + AI).}
\label{tab:results_ratio_right}
\resizebox{1.0\columnwidth}{!}{%
\begin{tabular}{llllll} \toprule
\multicolumn{1}{c}{\textbf{Ratio of `Right'}} &
  \multicolumn{1}{c}{\textbf{All}} &
  \multicolumn{1}{c}{\textbf{NoExp}} &
  \multicolumn{1}{c}{\textbf{Exp}} &
  \multicolumn{1}{c}{\textbf{TPs}} &
  \multicolumn{1}{c}{\textbf{NVs}} \\ \midrule
\begin{tabular}[c]{@{}l@{}}Human + AI\\ Numerical Confidence Scores (Condition A)\end{tabular} &
  66.14 &
  65.97 &
  66.43 &
  69.84 &
  64.29 \\ \midrule
\begin{tabular}[c]{@{}l@{}}Human + AI\\ Distance-based \\ Confidence Scores (Condition B)\end{tabular} &
  \textbf{74.34} &
  \textbf{74.79} &
  \textbf{73.57} &
  \textbf{77.78} &
  \textbf{72.62}  \\\midrule
\begin{tabular}[c]{@{}l@{}}Improvement\\(P-value)\end{tabular} &
  \begin{tabular}[c]{@{}l@{}}$\uparrow 8.20$\\ ($p<0.01$)\end{tabular} &
  \begin{tabular}[c]{@{}l@{}}$\uparrow 8.82$\\ ($p<0.01$)\end{tabular}  &
  \begin{tabular}[c]{@{}l@{}}$\uparrow 7.14$\\ ($p<0.01$)\end{tabular}  &
  \begin{tabular}[c]{@{}l@{}}$\uparrow 7.94$\\ ($p<0.01$)\end{tabular}  &
  \begin{tabular}[c]{@{}l@{}} $\uparrow 8.33$\\ ($p<0.01$)\end{tabular}  \\ 
  \bottomrule
\end{tabular}%
}
\end{table}

\subsubsection{Without vs With Exploring a Threshold of Confidence Scores}
Both \textbf{NonExp and Exp participants using Distance-based Confidence Scores with interactive example-based explanations achieved significantly higher rates of Right decisions compared to those using Numerical Confidence Scores}: 74.79\% vs. 65.97\% for NonExp (8.82\% increase, $p<0.01$) and 73.57\% vs. 66.4\% for Exp (7.14\% increase, $p<0.01$).

Further analysis of Right and Wrong decisions (Appendix. Table \ref{tab:results_ratio_right_wrong_detailed}) showed that using Distance-based Confidence Scores with interactive example-based explanations, {NoExp participants had a 7.15\% higher rate of agreeing with Right AI outputs} ($p<0.01$) and a {9.25\% lower rate of rejecting Right AI outputs} ($p<0.01$) than those using Numerical Confidence Scores; %\textbf{NoExp participants had a 7.15\% higher rate of agreeing with Right AI outputs} ($p<0.01$), a 1.68\% higher rate of rejecting Wrong AI outputs ($p=0.15$), a 0.42\% higher rate of agreeing with Wrong AI outputs ($p=0.41$), and a \textbf{9.25\% lower rate of rejecting Right AI outputs} ($p<0.01$) than those using Numerical Confidence Scores (Appendix. Table \ref{tab:results_ratio_right_wrong_detailed}); 
Also, {Exp participants had a 5.00\% higher rate of agreeing with Right AI outputs} ($p<0.01$) and a {9.29\% lower rate of rejecting Right AI outputs} ($p<0.01$) than those using Numerical Confidence Scores.
%Exp participants had a 4.25\% higher ratio of agreeing with `Right' AI outputs and a 4.80\% lower ratio of agreeing with `Wrong' AI outputs than `NoExp' 

\subsubsection{Experts vs Novices}
Among participant groups, domain experts, \textbf{TPs using Distance-based Confidence Scores with interactive explanations achieved the highest ratio of Right decisions, 7.94\% higher than those using Numerical Confidence Scores} ($p<0.01$). Similarly, novices, \textbf{NVs showed an 8.33\% improvement} with Distance-based Confidence Scores and interactive explanations over Numerical Confidence Scores ($p<0.01$). 

Further analysis of `Right' and `Wrong' decisions (Appendix. Table \ref{tab:results_ratio_right_wrong_detailed}) showed that using Distance-based Confidence Scores with interactive example-based explanations, {TPs had a 5.56\% higher rate of agreeing with Right AI outputs} ($p<0.01$) and a {7.15\% lower rate of rejecting Right AI outputs} ($p<0.01$) than those using Numerical Confidence Scores; Likewise, {NVs had a 6.74\% higher rate of agreeing with Right AI outputs} ($p<0.01$) and a {10.32\% lower rate of rejecting Right AI outputs} ($p<0.01$).

\subsection{Ratio of `Changed' decisions}
Overall, participants using Numerical Confidence Scores had lower ratios of `Changed' decisions than those using Distance-based Confidence Scores with example-based explanations across all participant groups except Exp participants (Appendix. Table \ref{tab:results_changed_decisions}). Specifically, participants using Numerical Confidence Scores (53.44\%) had a 5.03\% lower ratio of `Changed' decisions than those using Distance-based Confidence Scores (58.47\%) ($p=0.11$). In addition, using Distance-based Confidence Scores, NoExp participants (61.34\%) had a 10.92\% higher ratio ($p < 0.01$), Exp participants (53.57\%) had an 5.00\% lower ratio ($p < 0.05$), TPs (60.32\%) had a 10.32\% higher ratio ($p < 0.05$), NVs (57.54\%) had a 2.38\% higher ratio ($p=0.25$) than those using Numerical Confidence Scores.

\subsubsection{Ratio of `Right' decisions without/with AI }\label{sect:results_performance}
Overall, \textbf{participants using Distance-based Confidence Scores and example-based explanations achieved a complementary performance improvement of 3\% (from 71\% to 74\%) while those using Numerical Confidence Scores had a 1\% performance decline (from 67\% to 66\%)} (Appendix. Figure \ref{fig:results_performance}). 

This pattern of complementary performance gains was consistent across participant groups except for TPs. Using Distance-based Confidence Scores and interactive example-based explanations, NoExp participants improved by 3\% (from 71\% to 74\%); Exp participants improved by 3\% (from 70\% to 73\%); NVs improved by 6\% (from 66\% to 72\%). TPs using Distance-based Confidence Scores and interactive example-based explanations had a 3\% performance decline (from 80\% to 77\%) while TPs using Numerical Confidence Scores had 2\% performance decline (from 71\% to 69\%).

\subsection{Changed `Right' Decisions}\label{sect:results_changedright}
Overall, participants using Distance-based Confidence Scores and interactive example-based explanations had \textbf{a 40.48\% of `ChangedRight' decisions, which is a 7.15\% higher ratio than those using Numerical Confidence Scores (33.33\%) ($p<0.05$)} and {a 17.99\% of `ChangedWrong' decisions, which is 2.12\% lower than those using Numerical Confidence Scores (20.11\%) ($p = 0.24$)}.
Detailed ChangedRight decision ratios after reviewing AI outputs (Human + AI) by participant groups are provided in Appendix. Table \ref{tab:results_changed_decisions}.

%(Appendix. Table \ref{tab:results_changed_decisions}).

\subsubsection{With vs Without Exploring a Threshold of Confidence Scores}
 NoExp participants without exploring a threshold of confidence scores and using Distance-based Confidence Scores and interactive example-based explanations had a \textbf{42.44\% of ChangedRight decisions, which is 10.09\% higher than those using Numerical Confidence Scores} (32.35\%) ($p<0.01$) and {a 18.91\% of ChangedWrong decisions, which is 0.84\% lower than those using Numerical Confidence Scores (18.07\%) ($p=0.84$)}. 

Similarly, the Exp participants with exploring a threshold of confidence scores and using Distance-based Confidence Scores and example-based explanations had a 37.14\% of ChangedRight decisions, which is 2.14\% higher than those using Numerical Confidence Scores (35.00\%) ($p=0.29$) and \textbf{a 16.43\% of `ChangedWrong' decisions, which is 7.14\% lower than those using Numerical Confidence Scores} (23.57\%) ($p < 0.01$).

\subsubsection{Experts vs Novices}
Domain experts, TPs using Distance-based Confidence Scores and example-based explanations had a \textbf{43.65\% of ChangedRight decisions, which is 12.7\% higher than those using Numerical Confidence Scores} (30.95\%) ($p<0.01$) and {a 16.67\% of ChangedWrong decisions, which is 2.38\% lower than those using Numerical Confidence Scores} (19.05\%) ($p=0.25$). 
Similarly, novices (NVs) using Distance-based Confidence Scores and example-based explanations had a 38.89\% of ChangedRight decisions, 4.37\% higher than those using Numerical Confidence Scores ($p=0.11$) and {a 18.65\% of `ChangedWrong' decisions, 1.98\% lower than those using Numerical Confidence Scores (20.63\%) ($p = 0.22$)}.

\subsection{Duration of Decision Making}\label{sect:results_duration}
We estimated decision-making duration by measuring the time from when participants began reviewing a video to when they submitted an assessment score. Overall, Human + AI participants (553.77 secs) spent on average 101.18 seconds more on the tasks than Human  participants without reviewing AI outputs (452.59 secs). For \textbf{NoExp participants, while Human + AI participants (391.05 secs) had 82.36 seconds less on the assigned task than those in the Human-only condition (473.41 secs)} ($p<0.05$). In contrast, for Exp participants, Human + AI participants (716.49 secs) had 284.71 seconds more on the assigned tasks than those in Human (431.78 secs) ($p=0.07$).

%\subsection{Usability Responses}\label{sect:results_usability}

\subsection{Post-Study Questionnaires}
Among 27 participants, 11 preferred the system with Numerical Confidence Scores (6 totally; 3 much more; 2 slightly more), 7 were neutral, and 9 preferred the system with Distance-based Confidence Scores and example-based explanations (1 totally; 6 much more; 2 slightly more). 
Some participants preferred to use the Numerical Confidence Scores system as it is ``\textit{faster and easier to read}'' (TP9). They expressed the preference to review confidence scores in ``\textit{a value instead of a distance-based value}'' 
(NV07), which helped participants to ``\textit{brought[bring] my attention to low confidence scores}'' (NV1). In contrast, others preferred the system with Distance-based Confidence Scores and interactive example-based explanations as it provided ``\textit{more informative}'' (TP7) and ``\textit{more accurate than a numerical confidence score}'' (NV17). In addition, participants found that the system with Distance-based Confidence Scores and interactive explanations 
%``\textit{detailed information}'' (NV17) 
is useful support to ``\textit{verify by showing a few cases with dots (embedding data)''} (NV9) and including photos of other post-stroke survivors, relevant examples (TP7).

Domain experts (TPs) and novices (NVs) differed in their rankings of influential system components (Appendix. Table \ref{tab:overall_questions}). TPs rated `Exploring confidence scores for task delegation' and `Task delegation to AI' higher while they rated example-based explanations more useful than feature analysis. In contrast, NVs ranked 'Numerical confidence scores by AI' and feature analysis as the most useful components of the system.

\section{Discussion}
%Achieving appropriate reliance on AI \cite{buccinca2021trust,lee2023understanding,salimzadeh2024dealing} is essential for effective human-AI collaborative decision-making. 
In this section, we discuss how exploring confidence score thresholds for task delegation to AI and utilizing distance-based confidence scores with interactive example-based explanations influence users' reliance on AI. Also, we highlight ongoing challenges for human-AI collaborative decision-making.

\subsection{Effectiveness of Interactive Explorations of Confidence Scores \& Task Delegation}
Our interactive exploration of confidence score thresholds (Exp) assisted participants to have a 3.58\% lower ratio of agreeing with Wrong AI outputs (overreliance) using Numerical Confidence Scores, and a 1.85\% lower rate using Distance-based Confidence Scores and interactive example-based explanations than those without exploring confidence score thresholds (Appendix. Table \ref{tab:results_ratio_right_wrong_detailed}). 

For decision improvement, {Exp participants demonstrated a 2.65\% higher ratio of ChangedRight decisions} using Numerical Confidence Scores and {a 0.59\% higher ratio of Right decisions} using Distance-based Confidence Scores and interactive example-based explanations than NoExp participants, indicating enhanced reliance on AI after interactive explorations of confidence score thresholds. 

Both Exp and NoExp participants using Distance-based Confidence Scores and example-based explanations achieved significantly higher ratios of Right decisions - 8.82\% higher and 7.14\% improvements respectively ($p<0.01$) (Table \ref{tab:results_ratio_right}) and complementary enhancement in performance. This contrasts with Exp and NoExp participants using Numerical Confidence Scores, who exhibited performance declines (Appendix. Figure \ref{fig:results_performance}). Specifically, using Distance-based Confidence Scores, Exp participants had a 7.14\% lower ratio of ChangedWrong decisions ($p<0.01$) and NoExp participants showed a 10.09\% higher ratio of ChangedRight decisions ($p<0.01$) than those using Numerical Confidence Scores. 

In contrast to previous studies \cite{buccinca2021trust,lee2023understanding} that describe users' overreliance on AI after reviewing important feature explanations, our system introduced additional functionalities of exploring confidence score thresholds (Figure \ref{fig:interface_expconf}) and task delegation to AI (Figure \ref{fig:interface_taskdelegate}). Our study implies that users can have greater understanding and control through explorations of confidence scores to foster more analytical review of AI even with important features \cite{buccinca2021trust,lee2023understanding} or numerical confidence values \cite{zhang2020effect,prabhudesai2023understanding}.

Similar to prior studies describing improved task performance and task satisfaction in workplaces through AI delegation \cite{hemmer2023human}, our study also demonstrated the potential of task delegation to AI to enhance decision-making performance and practical efficiency. Specifically, NoExp participants completed their tasks 82.36 seconds faster on average than those in the Human only condition without AI assistance (Section \ref{sect:results_duration}). 

%\subsection{`Numerical Confidence Scores' vs `Distance-based Confidence Scores'}
\subsection{Numerical vs Distance-based Visualization of Confidence Scores}
Our Distance-based Confidence Scores and interactive example-based explanations were consistently more effective in enhancing users' decision-making across all participant groups than Numerical Confidence Scores: improvement of 8.2\% for all, 8.82\% for NoExp participants, 7.14\% for Exp participants, 7.94\% for experts, and 8.33\% for novices ($p<0.01$) (Table \ref{sect:results_changedright}). A detailed analysis showed that these improvements stemmed from both a higher agreement with correct AI outputs and a lower rejection of correct AI outputs across all groups ($p<0.01$) (Appendix. Table \ref{tab:results_ratio_right_wrong_detailed}). 

In addition, participants using Distance-based Confidence Scores exhibited a significantly higher rate of ChangedRight decisions for all participant groups except Exp participants and significantly lower ChangedWrong decisions specifically for Exp participants. These results highlight the effectiveness of Distance-based Confidence Scores and interactive example-based explanations in fostering more accurate and calibrated reliance on AI outputs.

Participants using Numerical Confidence Scores and important feature explanations had a performance decline similar to prior studies reporting overreliance on AI \cite{buccinca2021trust,lee2023understanding}. In contrast, participants using Distance-based Confidence Scores and interactive example-based explanations demonstrated complementary performance improvement and significantly higher rates of ChangedRight decisions for all, NoExp participants, and domain experts ($p<0.01$) as well as a significantly lower rate of ChangedWrong decisions for Exp participants ($p<0.01$). These results suggest that Distance-based Confidence Scores and interactive example-based explanations support more analytical review of AI outputs and foster more appropriate reliance on AI compared to Numerical Confidence Scores.

Both domain experts and novices using Distance-based Confidence Scores and interactive example-based explanations had similar trends in achieving higher ratios of Right decisions. However, novices showed greater improvement, suggesting that they were more influenced by the AI outputs. 

Post-study responses also showed differing preferences by domain experts and novices. While domain experts appreciated the value of exploring confidence score thresholds for task delegation, novices preferred reviewing AI outputs including confidence scores and feature explanations to build familiarity and confidence in performing the task. Unlike prior research that reports improved task satisfaction regardless of AI delegation \cite{hemmer2023human}, our findings highlight the importance of tailoring AI delegation mechanisms by characterizing user profiles and contexts \cite{suresh2021beyond}. Designing systems that adapt to users’ experience levels and decision-making styles is essential to supporting more effective and trustworthy human-AI collaborative decision-making.

For effective human-AI collaborative decision-making, our study highlights ongoing challenges in educating users about ML components (e.g. AI outputs, confidence scores) \cite{cai2019hello,lee2024improving} and effectively communicating this information \cite{zhang2020effect,prabhudesai2023understanding} to foster informed, analytical engagement. Our work emphasizes the importance of having an appropriate balance between encouraging critical review of AI outputs while minimizing cognitive or interactional burdens, particularly for users without technical backgrounds.

\subsection{Limitations}
As this work primarily focuses on examining the effect of interactive explorations and distance-based visualizations of confidence scores with example-based explanations, our study does not investigate how other forms of AI explanations or other forms of uncertainty visualizations \cite{hullman2018pursuit} may influence users' reliance on AI during human-AI collaborative decision-making. In addition, our work is limited to exploring only a feed-forward neural network using video-based input data for physical stroke rehabilitation assessment. While a small sample size of participants is not unusual in similar previous work of human-AI collaborative decision-making in health \cite{lee2020co,bussone2015role}, our participants do not fully represent the broader population of health professionals. Thus, further research is needed with diverse participant groups, a wider range of decision-making tasks, other types of AI/ML models, data modalities, and visualizations to enhance the generalization of our findings.

\section{Conclusion}
In this work, we contributed to an empirical study to understand the effect of interactive exploration of confidence scores for task delegation to AI and distance-based visualization with interactive example-based explanations on users' reliance on AI for human-AI collaborative decision-making. Our findings showed that distance-based uncertainty scores outperformed traditional probability-based uncertainty scores in identifying uncertain cases. In addition, our study with domain experts and novices (i.e. other health professionals and students in medicine, nursing, and therapy) showed that interactive explorations and visualizations of confidence scores enabled users to achieve significantly more correct decisions while assisting them to change their decisions to correct ones and reducing their wrong changes in decisions after reviewing AI outputs. These findings highlight the potential of interactive explanations and functionalities to foster analytical reviews of AI/ML components (i.e. confidence scores) for improving AI-assisted decision-making. In addition, our work emphasizes ongoing challenges in educating participants without AI backgrounds about these AI/ML components and achieving an appropriate balance between encouraging critical reviews of AI outputs while minimizing their burdens for effective human-AI collaborative decision-making.

\begin{acks}
The authors would like to thank all the participants for their time and valuable inputs to this work. We also thank the anonymous reviewers for their constructive feedback. This research is supported by the Ministry of Education, Singapore under its Academic Research Fund Tier 2 (MOE-T2EP20223-0007). Any opinions, findings and conclusions or recommendations expressed in this material are those of the authors and do not reflect the views of the Ministry of Education, Singapore.
\end{acks}

\bibliographystyle{ACM-Reference-Format}
\bibliography{main}

%\appendix
\appendix

\section{System Implementation Details}
\subsection{Feature Extractions}\label{appendix:feature_extract}
We built upon the previous research on rehabilitation assessment \cite{lee2019learning} to process the estimated joint positions of post-stroke survivors' exercises and extract various kinematic features. 
For the kinematic features of the `Range of Motion' (ROM), we computed joint angles, such as elbow flexion, shoulder flexion, and elbow extension, and normalized relative trajectory (i.e. the Euclidean distance between two joints - head and wrist; head and elbow), and the normalized trajectory distance (i.e. the absolute distance between two joints - head and wrist, shoulder and wrist) in the x, y, and z coordinates \cite{lee2019learning}. For the features of the `Compensation', we calculated the normalized trajectories, which indicate the distances between joint positions of the head, spine, and shoulder in the x, y, and z coordinates from the initial to the current frame over the entire exercise motion \cite{lee2019learning}.

\subsection{Results of Machine Learning Models}\label{appendix:results_ml}

For the feed-forward NN model, we performed a grid-search over various architectures (i.e. one to four layers with 32, 64, 128, 256, 512 hidden units) and learning rates (i.e. 0.00001, 0.0005, 0.0001, 0.005, 0.001, 0.01) and minimized the cross entropy loss to train the model. %with respect to $\textbf{W}$. 
To train and evaluate the model, we utilized the leave-one-subject-out cross-validation, where we trained the model with data from all post-stroke survivors and healthy participants except one post-stroke survivor or healthy participant and tested the model with data from the held-out post-stroke survivor or participant. 

\begin{table}[htp]
\centering
\caption{Results of Machine Learning Models to Assess Quality of Motion (Feed-Forward Neural Network, MCDropout Network, RBF Network) and Predict True Class Probability for Uncertainty Quantification (Confidence Network)}
\label{tab:results_mlmodels}
\resizebox{1.0\columnwidth}{!}{%
\begin{tabular}{clcccccc} \toprule
\multicolumn{2}{c}{\textbf{Data}} &
  \multicolumn{3}{c}{\textbf{All (Post-Stroke + Healthy)}} &
  \multicolumn{3}{c}{\textbf{Post-Stroke}} \\
Models &
  Labels &
  ROM &
  COMP &
  Overall &
  ROM &
  COMP &
  Overall \\ \midrule
\begin{tabular}[c]{@{}c@{}}Feed-Forward\\ NN\end{tabular} &
  \begin{tabular}[c]{@{}l@{}}Assessment\\ Scores\end{tabular} &
  77.97 &
  85.97 &
  \textbf{81.97} &
  82.73 &
  82.82 &
  \textbf{82.76} \\ \midrule
\begin{tabular}[c]{@{}c@{}}Confidence\\ Network\end{tabular} &
  \begin{tabular}[c]{@{}l@{}}True Class\\ Probability\end{tabular} &
  0.0905 &
  0.0605 &
  0.0755 &
  0.1110 &
  0.0624 &
  0.0867 \\ \midrule
\begin{tabular}[c]{@{}c@{}}MCDrop\\ Network\end{tabular} &
  \begin{tabular}[c]{@{}l@{}}Assessment\\ Scores\end{tabular} &
  75.67 &
  84.67 &
  80.17 &
  81.73 &
  81.98 &
  81.85 \\ \midrule
\multicolumn{1}{l}{RBF Network} &
  \begin{tabular}[c]{@{}l@{}}Assessment\\ Scores\end{tabular} &
  75.67 &
  65.47 &
  70.57 &
  81.79 &
  72.84 &
  77.29 \\ \bottomrule
\end{tabular} 
} 
\end{table}

\subsection{Implementation Details of Uncertainty Quantification Approaches}\label{appendix:uq}

\textbf{Confidence Estimation Approach}: We followed the work that proposed to utilize True Class Probability (TCP) associated to the true class $Y*$ and learn a model to estimate this TCP in the context of predicting a model failure \cite{corbiere2019addressing}. When a model makes incorrect predictions, the probability associated with the true class $Y^*$ is likely to have a low value that indicates an erroneous prediction. We define the TCP as follows: 
%$TCP(\textbf{X}, Y^*) = P(Y = Y^* |\textbf{W, X}) = c^*(\textbf{X}, Y*)$
\begin{equation}
TCP(\textbf{X}, Y^*) = P(Y = Y^* |\textbf{W, X}) = c^*(\textbf{X}, Y*)
\end{equation}

As the true class $Y^*$ is not available during testing, we aim to learn a model called \textit{`Confidence Network'} with parameters $\theta$ that predicts a TCP confidence score $c^*(\textbf{X}, Y^*) = P(Y = Y^* |\textbf{W, X})$. During training a confidence network, we aim to seek the parameters of a network $\theta$ that $\hat{c}(\textbf{X}, \theta)$ makes as close to $c^*(\textbf{X}, Y^*)$ using the following $l_2$ loss:
%$\mathcal{L}_{conf}^{}= E(\hat{c}, c^*) = \frac{1}{N}\sum_{i=1}^N (\hat{c}(\textbf{X}_i, \theta) - c^*(\textbf{X}_i, Y_i^*))^2)$
\begin{equation}
\mathcal{L}_{conf}^{}= E(\hat{c}, c^*) = \frac{1}{N}\sum_{i=1}^N (\hat{c}(\textbf{X}_i, \theta) - c^*(\textbf{X}_i, Y_i^*))^2)
\end{equation}

For the implementations of a Confidence Network, we grid-searched various architectures (i.e. one to three layers with 32, 64, 128, 256, 512 hidden units) and learning rates (i.e. 0.00001, 0.0005, 0.0001, 0.005, 0.001, 0.01). Similar to the implementation of a feed-forward NN model, we applied the leave-one-subject-out cross-validation to train and evaluate the Confidence Network.

The final model architectures and learning rates of Confidence Networks are three layers of 32 hidden units and a learning rate of 0.005 for the ROM performance component and three layers of 256 hidden units and a learning rate of 0.01. Overall, the Confidence Networks achieved an average of 0.0755 mean square error (MSE) to replicate the TCP scores of the `ROM' and `Compensation' performance components using data from post-stroke survivors and healthy participants: an average of 0.0905 MSE for the `ROM' performance component and an average of 0.0605 MSE for the `Compensation performance component (Appendix. Table \ref{tab:results_mlmodels}). For the post-stroke survivors' data, the Confidence Networks achieved an average of 0.0867 MSE to replicate the TCP scores of the `ROM' and `Compensation' performance components: an average of 0.1110 MSE for the `ROM' performance component and an average of 0.0624 MSE for `Compensation' performance component (Appendix. Table \ref{tab:results_mlmodels}).

\textbf{Bayesian approaches} for uncertainty quantification of neural networks also have gained a lot of attention. Specifically, we built upon the work by Gal and Ghahramani that proposed to use Monte Carlo Dropout (MCDropout) to estimate the posterior distribution of model predictions using several stochastic network predictions \cite{gal2016theoretically}. We estimate the posterior predictive distribution by sampling several stochastic network predictions \cite{gal2016theoretically}:

\begin{equation}
\begin{split}
 \hat{c} (Y = c|\textbf{X}, \mathcal{D}_{train} ) &= \int p(Y = c|\textbf{X, W})P(\textbf{W}|  \mathcal{D}_{train}) d\textbf{W}\\
 & \approx \frac{1}{T}\sum_{t=1}^{T}P(Y|\textbf{X}, \textbf{$W^{(t)}$})
\end{split}
\end{equation}

For the implementations of an MCDropout Network, we followed the architectures of our feed-forward NN models and explored various dropout rates from 0.1 to 0.5 using the leave-one-subject-out cross-validation. The final dropout probability was specified as 0.3 for both ROM and Compensation performance components. Overall, the MCDropout Network performed an average of 80.17\% F1-score to replicate therapist's assessments using data from post-stroke survivors and healthy participants: an average of 75.67\% F1-score for the `ROM' performance component and an average of 84.67\% F1-score for the `Compensation' performance component (Appendix. Table \ref{tab:results_mlmodels}). Using only post-stroke survivors' data, the MCDropout Network performed an average of 81.85\% F1-score to replicate therapist's assessments: an average of 81.73\% F1-score for the `ROM' performance component and an average of 81.98\% F1-score for the `Compensation' performance component (Appendix. Table \ref{tab:results_mlmodels}). 

\textbf{Distance-based approaches} for uncertainty quantification of neural networks is to utilize the input or vector space of a neural network and compute the distances in the trust score matters \cite{jiang2018trust,papernot2018deep,van2020uncertainty}. Specifically, we explored NN-based and radial basis function (RBF)-based distance approaches. The \textbf{NN-based} distance approach is to explore the intermediate representations of a feed-forward NN model to compute the centroid of classes and input data \cite{jiang2018trust,papernot2018deep}. The \textbf{RBF-based} distance approach is to implement an RBF network with a set of feature vectors corresponding to the different classes (centroids) \cite{van2020uncertainty}.  After computing the centroids of classes and input data using either an NN-based or an RBF-based model, we then utilized the distance between the model output and the closest centroid as the uncertainty \cite{jiang2018trust,papernot2018deep,van2020uncertainty}. 

For the RBF-based Networks, we applied the leave-one-subject-out cross-validation and grid-searched various architectures: one or two layers of radial basis functions (RBF) with centroids (i.e. 30, 25, 20, 15, 10, 5, 3) and hidden units (i.e. 16, 32, 64, 128, 256, 512) and learning rates (i.e. 0.00001, 0.0005, 0.001, 0.01, 0.1). The final model architectures and learning rates of RBF-based Networks are one RBF layer of 15 centroids, one layer of 512 hidden units, and one RBF layer of 3 centroids with the learning rate of 0.001 for the `ROM' performance component; one RBF layer of 3 centroids, one layer of 16 hidden units, and the learning rate of 0.01 for the `Compensation' performance component. Overall, the RBF Network performed an average of 70.57\% F1-score to replicate therapist's assessments using data from post-stroke survivors and healthy participants: an average of 75.67\% F1-score for the `ROM' performance component and an average of 65.47\% F1-score for the `Compensation' performance component (Appendix. Table \ref{tab:results_mlmodels}). For post-stroke survivors' data, the RBF Network performed an average of 77.29\% F1-score to replicate the therapist's assessments: an average of 81.79\% F1-score for the `ROM' performance component and an average of 77.29\% F1-score for the `Compensation' performance component (Appendix. Table \ref{tab:results_mlmodels}).

% \usepackage{graphicx}
\begin{table}[htp]
\centering
\caption{Results of Identifying Uncertain Cases using Machine Learning Models and Uncertainty Quantification Techniques. The feed-forward Neural Network models using a distance-based uncertainty quantification approach achieved the highest performance}
\label{tab:results_uqmodels}
\resizebox{1.0\columnwidth}{!}{%
\begin{tabular}{ccccccccccc} \toprule
\multicolumn{2}{c}{\textbf{}} &
  \multicolumn{3}{c}{\textbf{All (Post-Stroke + Healthy)}} &
  \multicolumn{3}{c}{\textbf{Post-Stroke}} &
  \multicolumn{3}{c}{\textbf{Healthy}} \\ 
\multicolumn{2}{c}{\textbf{UQ Models}}            & ROM   & COMP  & Overall & ROM   & COMP  & Overall & ROM   & COMP  & Overall \\ \midrule
\begin{tabular}[c]{@{}c@{}}Feed-Forward\\ NN \cite{hendrycks2016baseline} \end{tabular} &
  \begin{tabular}[c]{@{}c@{}}Maximum Class\\ Probability (MCP)\end{tabular} &
  77.31 &
  88.41 &
  82.86 &
  84.49 &
  86.18 &
  85.33 &
  66.54 &
  91.75 &
  79.14 \\ \midrule
\begin{tabular}[c]{@{}c@{}}Confidence\\ Network \cite{corbiere2019addressing}\end{tabular} &
  \begin{tabular}[c]{@{}c@{}}Estimating\\ True Class\\ Probability\end{tabular} &
  76.74 &
  83.15 &
  79.94 &
  84.82 &
  81.45 &
  83.13 &
  64.63 &
  85.71 &
  75.17 \\ \midrule
\begin{tabular}[c]{@{}c@{}}MCDrop\\ Network \cite{gal2016theoretically}\end{tabular} & Bayesian       & 58.71 & 87.13 & 72.92   & 72.96 & 84.64 & 78.80   & 37.33 & 90.87 & 64.10   \\ \midrule
RBF Network \cite{van2020uncertainty}                                             & Distance-based & 75.47 & 78.39 & 76.93   & 86.67 & 85.68 & 86.17   & 58.67 & 62.08 & 60.37   \\ \midrule
\begin{tabular}[c]{@{}c@{}}Feed-Forward\\ NN\end{tabular} &
  Distance-based &
  \textbf{89.21} &
  \textbf{94.14} &
  \textbf{93.63} &
  \textbf{90.29} &
  \textbf{93.26} &
  \textbf{91.77} &
  \textbf{87.60} &
  \textbf{95.46} &
  \textbf{91.53} \\ \bottomrule
\end{tabular}%
}
\end{table}

\subsection{Results of Dimensional Reduction Techniques}\label{appendix:results_drt}

% \usepackage{graphicx}
\begin{table}[h]
\centering
\caption{Results of k-Nearest Neighbor Classifiers using Dimensional Reduction Techniques}
\label{tab:results-drt}
\resizebox{\columnwidth}{!}{%
\begin{tabular}{ccccccccc} \toprule
     & \begin{tabular}[c]{@{}c@{}}Distance\\ Metric\end{tabular} & K=5   & K=10  & K=15  & K=20  & K=30  & Avg            & Max            \\ \midrule
UMAP & Euclidean & 76.35 & 77.50 & 76.91 & 76.25 & 77.33 & 76.87 & 77.50 \\ \midrule
UMAP & Cosine    & 59.67 & 64.60 & 64.01 & 63.37 & 65.40 & 63.41 & 65.40 \\ \midrule
PCA  & Euclidean & 77.67 & 78.23 & 76.90 & 77.43 & 74.25 & 76.90 & 78.23 \\ \midrule
PCA  & Cosine    & 72.92 & 74.61 & 75.48 & 75.11 & 75.83 & 74.79 & 75.83 \\ \midrule
t-SNE & Euclidean                                                 & 79.65 & 80.54 & 79.13 & 80.15 & 77.87 & \textbf{79.47} & \textbf{80.54} \\ \midrule
t-SNE & Cosine    & 69.73 & 70.55 & 69.09 & 70.14 & 71.74 & 70.25 & 71.74 \\ \bottomrule
\end{tabular}%
}
\end{table}

%\newpage
\section{Details of User Study}
% \usepackage{graphicx}
% \usepackage{graphicx}
\begin{table*}[htp]
\caption{Detailed Demographics of Participants: Domain experts, therapists, who have experience in stroke rehabilitation more than one year (TP1 - TP9) and novices (NV1 - NV17) (i.e. other health professionals and students majoring in medicine, therapy)}
\label{tab:participants_details}
\resizebox{\textwidth}{!}{%
\begin{tabular}{lllllrl} \toprule
PID  & Sex    & Age           & Occupation                                & Setting                     & \multicolumn{1}{l}{\# of yrs} & Condition \\  \midrule
TP1  & Female & 25 - 34 years & Occupational Therapist (OT)               & Inpatient Rehabilitation    & 12                            & noExplore \\
TP2  & Female & 25 - 34 years & Occupational Therapist (OT)               & Acute Care                  & 3                             & noExplore \\
TP3  & Male   & 25 - 34 years & Physiotherapist                           & Acute Care                  & 1.25                          & explore   \\
TP4  & Male   & 25 - 34 years & PhysioTherapist (PT)                      & Inpatient Rehabilitation    & 7                             & explore   \\
TP5  & Female & 45 - 54 years & Occupational Therapist (OT)               & Acute Care                  & 26                            & noExplore \\
TP6  & Female & 25 - 34 years & Occupational Therapist (OT)               & Inpatient Rehabilitation    & 6                             & noExplore \\
TP7  & Female & 35 - 44 years & PhysioTherapist (PT)                      & Community Rehab             & 18                            & noExplore \\
TP8  & Female & 25 - 34 years & Occupational Therapist (OT)               & Skilled Nursing Facility    & 5                             & noExplore \\
TP9  & Female & 35 - 44 years & Occupational Therapist (OT)               & Outpatient                  & 15                            & explore   \\
NV01 & Female & 18 - 24 years & Physiotherapy Assistant                   & Internal Medicine, Geriatic & 0.5                           & explore   \\
NV02 & Female & 25 - 34 years & Older Adult Health care associate         & Nursing Facility            & 2                             & noExplore \\
NV03 & Male   & 25 - 34 years & Health Assistant                          & Inpatient Geriatrics        & 2                             & noExplore \\
NV04 & Female & 25 - 34 years & Healthcare for Inpatient Medicine         & Inpatient                   & 5                             & noExplore \\
NV05 & Male   & 25 - 34 years & Health Assistant for Inpatient Geriatrics & Inpatient                   & 2                             & noExplore \\
NV06 & Male   & 25 - 34 years & Student in Physiotherapy                  &                             & \multicolumn{1}{l}{}          & noExplore \\
NV07 & Male   & 18 - 24 years & Student in Physiotherapy                  &                             & \multicolumn{1}{l}{}          & noExplore \\
NV08 & Male   & 35 - 44 years & Student in Physiotherapy                  &                             & \multicolumn{1}{l}{}          & explore   \\
NV09 & Male   & 18 - 24 years & Student in Medicine                       &                             & \multicolumn{1}{l}{}          & noExplore \\
NV10 & Female & 18 - 24 years & Student in Medicine                       &                             & \multicolumn{1}{l}{}          & noExplore \\
NV11 & Male   & 18 - 24 years & Student in Nursing                        &                             & \multicolumn{1}{l}{}          & noExplore \\
NV12 & Female & 35 - 44 years & Student in Nursing                        &                             & \multicolumn{1}{l}{}          & explore   \\
NV13 & Female & 18 - 24 years & Student in Nursing                        &                             & \multicolumn{1}{l}{}          & explore   \\
NV14 & Female & 18 - 24 years & Student in Healthcare                     &                             & \multicolumn{1}{l}{}          & explore   \\
NV15 & Female & 18 - 24 years & Respite caregiver for patients with dementia and post-stroke survivors & Inpatient/acute and home care & 2 & noExplore \\
NV16 & Female & 25 - 34 years & Nurse                                     & Outpatient                  & 7                             & noExplore \\
NV17 & Male   & 25 - 34 years & Nurse                                     & Acute/Inpatient Setting     & 6                             & explore  \\ 
NV18 & Male   & 25 - 34 years & Social Worker                                     &    &                              & explore  \\
NV19 & Male   & 18 - 24 years & Student in Medicine                                     &      &                              & explore  \\ 
NV20 & Male   & 35 - 54 years & Doctor                                   &      &                              & noExplore  \\ 
\bottomrule
\end{tabular}%
}
\end{table*}

\subsection{Data Analysis Metrics}\label{appendix:study_metrics}
\textbf{Ratio of `Right' Decisions:}
We computed the ratio of `Right' decisions by the participants, using the annotations of a therapist, who conducted the clinically validated functional assessment test, as ground truths. In addition, we further analyzed the decisions by the participants into (1) agreeing with `Right' AI outputs, (2) rejecting `Wrong' AI outputs, (3) agreeing with `Wrong' AI outputs (i.e. overreliance), and (4) rejecting `Right' AI outputs (i.e. underreliance) for detailed analysis.
%\cite{lai2021towards,lee2023understanding}. 

\textbf{Ratio of `Changed' Decisions:}
We also computed the ratio of `Changed' decisions by the participants after exploring a confidence score and reviewing AI outputs and explanations of our system to understand the effect of our system.
%\cite{lee2023understanding}.

\textbf{Ratio of Right Decisions before/after reviewing AI outputs}
Another metric is to compute the participants' performance on decision-making tasks before and after exploring a confidence score and reviewing AI outputs and explanations of our system.
%\cite{lai2021towards,lee2023understanding}.

\textbf{Ratio of `ChangedRight' Decisions:}
In addition, we further analyzed the ratio of how much participants changed their decisions into `Right' decisions (`ChangedRight') after exploring a confidence score and reviewing AI outputs and explanations of our system.

%\textbf{5) Duration of Decision-Making:}
%Our systems computed the estimated duration of decision-making by measuring the time from reviewing a video for the assessment to submit an assessment score.

\textbf{Post-Study Questionnaires:} 
At the end of our study, we asked participants to rate on a 7-point scale which of the following components \textit{``helped me validate the competence of the system''} and \textit{``determine the reliance on the system''}:
(a) introduction and tutorial materials, (b) a video of the patient's exercises, (c) exploring a threshold of confidence scores (Figure \ref{fig:interface_expconf}), (d) task delegation to AI (Figure \ref{fig:interface_taskdelegate}), (e) a predicted score by AI, (f) a numerical confidence score by AI, (g) a distance-based confidence score by AI, (h) important feature explanation, (i) relevant information of an example-based explanation, and (j) relevant images of an example-based explanation.

\newpage
\section{Details of User Study Results}

\begin{table*}[htp]
\centering
\caption{Detailed Ratio of `Right' and 'Wrong' decisions after reviewing AI outputs (`Human + AI') by all participants (All), participants who did not explore confidence scores before task delegations using confidence scores (NoExp), participants who explore confidence scores before task delegation (Exp), domain experts, therapists (TPs), and other health professionals and students (NVs)}
\label{tab:results_ratio_right_wrong_detailed}
\resizebox{1.0\textwidth}{!}{%
\begin{tabular}{lccccccccccccccc} \toprule
\multicolumn{1}{c}{\textbf{}} &
  \multicolumn{3}{c}{\textbf{All}} &
  \multicolumn{3}{c}{\textbf{NoExp}} &
  \multicolumn{3}{c}{\textbf{Exp}} &
  \multicolumn{3}{c}{\textbf{TPs}} &
  \multicolumn{3}{c}{\textbf{NVs}} \\ 
 &
  \textbf{Cond A} &
  \textbf{Cond b} &
  \textbf{P-Value} &
  \textbf{Cond A} &
  \textbf{Cond b} &
  \textbf{P-Value} &
  \textbf{Cond A} &
  \textbf{Cond b} &
  \textbf{P-Value} &
  \textbf{Cond A} &
  \textbf{Cond b} &
  \textbf{P-Value} &
  \textbf{Cond A} &
  \textbf{Cond b} &
  \textbf{P-Value} \\ \midrule
\begin{tabular}[c]{@{}l@{}}Right Decisions\end{tabular} &
  66.14 &
  \textbf{74.34} &
  {\textbf{0.0002}} &
  65.97 &
  \textbf{74.79} &
  \textbf{0.007} &
  66.43 &
  \textbf{73.57} &
  \textbf{0.001} &
  69.84 &
  \textbf{77.78} &
  \textbf{0.004} &
  64.29 &
  \textbf{72.62} &
  \textbf{0.004} \\ \midrule
\begin{tabular}[c]{@{}l@{}}Right Decisions - \\ Agree with \\ 'Right' AI outputs\end{tabular} &
  56.61 &
  \textbf{62.96} &
  {\textbf{0.0007}} &
  55.88 &
  \textbf{63.03} &
  \textbf{0.008} &
  57.86 &
  \textbf{62.86} &
  \textbf{0.00002} &
  59.52 &
  \textbf{65.08} &
  \textbf{0.008} &
  55.16 &
  \textbf{61.9} &
  \textbf{0.001} \\ \midrule
\begin{tabular}[c]{@{}l@{}}Right  Decisions - \\ Reject \\ 'Wrong' AI Outputs\end{tabular} &
  9.52 &
  11.38 &
  {0.14} &
  10.08 &
  {11.76} &
  {0.15} &
  8.57 &
  10.71 &
  0.24 &
  10.32 &
  12.7 &
  0.08 &
  9.13 &
  10.71 &
  0.17 \\ \midrule
\begin{tabular}[c]{@{}l@{}}Wrong  Decisions\end{tabular} &
  33.86 &
  \textbf{25.66} &
  \textbf{0.0009} &
  34.03 &
  \textbf{25.21} &
  \textbf{0.0007} &
  33.57 &
  \textbf{26.43} &
  \textbf{0.001} &
  30.16 &
  \textbf{22.22} &
  \textbf{0.004} &
  35.71 &
  \textbf{27.38} &
  \textbf{0.0003} \\ \midrule
\begin{tabular}[c]{@{}l@{}}Wrong Decisions -\\ Agree with\\ 'Wrong' AI outputs\end{tabular} &
  12.96 &
  14.02 &
  {0.27} &
  14.29 &
  {14.71} &
  {0.41} &
  10.71 &
  12.86 &
  0.27 &
  13.49 &
  12.7 &
  0.31 &
  12.7 &
  14.68 &
  0.19 \\ \midrule
\begin{tabular}[c]{@{}l@{}}Wrong Decisions - \\ Reject\\ 'Right' AI outputs\end{tabular} &
  20.9 &
  \textbf{11.64} &
  \textbf{0.00004} &
  19.75 &
  \textbf{10.5} &
  \textbf{0.0002} &
  22.86 &
  \textbf{13.57} &
  \textbf{\textless 0.00001} &
  16.67 &
  \textbf{9.52} &
  \textbf{0.002} &
  23.02 &
  \textbf{12.7} &
  \textbf{\textless .00001} \\ \bottomrule
\end{tabular}%
}
\end{table*}

\begin{table*}[htp]
\centering
\caption{Ratio of 'Changed', `ChangedRight', and 'ChangedWrong' decisions after reviewing AI outputs (`Human + AI') by all participants (All), participants without exploring a threshold of confidence scores (NoExp), participants with exploring a threshold of confidence scores (Exp), domain experts, therapists (TPs), and health professionals and students (NVs)}
\label{tab:results_changed_decisions}
\resizebox{\textwidth}{!}{%
\begin{tabular}{lccccccccccccccc} \toprule
\multicolumn{1}{c}{\textbf{}} &
  \multicolumn{3}{c}{\textbf{All}} &
  \multicolumn{3}{c}{\textbf{NoExp}} &
  \multicolumn{3}{c}{\textbf{Exp}} &
  \multicolumn{3}{c}{\textbf{TPs}} &
  \multicolumn{3}{c}{\textbf{NVs}} \\ \midrule
 &
  \textbf{CondA} &
  CondB &
  {\textbf{P-Value}} &
  CondA &
  CondB &
  \textbf{P-Value} &
  CondA &
  CondB &
  \textbf{P-Value} &
  CondA &
  CondB &
  \textbf{P-Value} &
  CondA &
  CondB &
  \textbf{P-Value} \\ \midrule
\begin{tabular}[c]{@{}l@{}}'Changed' Decisions\\ Overall\end{tabular} &
  53.44 &
  58.47 &
  {0.11} &
  50.42 &
  \textbf{61.34} &
  \textbf{0.009} &
  58.57 &
  \textbf{53.57} &
  \textbf{0.04} &
  50.00 &
  \textbf{60.32} &
  \textbf{0.028} &
  55.16 &
  57.54 &
  0.25 \\ \midrule
\begin{tabular}[c]{@{}l@{}}'ChangedRight' \\ Decisions\end{tabular} &
  33.33 &
  \textbf{40.48} &
  {\textbf{0.03}} &
  32.35 &
  \textbf{42.44} &
  \textbf{0.006} &
  35.00 &
  37.14 &
  0.29 &
  30.95 &
  \textbf{43.65} &
  \textbf{0.003} &
  34.52 &
  38.89 &
  0.11 \\ \midrule
\begin{tabular}[c]{@{}l@{}}'ChangedWrong' \\ Decisions\end{tabular} &
  20.11 &
  17.99 &
  {0.24} &
  18.07 &
  {18.91} &
  {0.84} &
  23.57 &
  \textbf{16.43} &
  \textbf{0.0006} &
  19.05 &
  16.67 &
  0.25 &
  20.63 &
  18.65 &
  {0.22} \\ \bottomrule
\end{tabular}%
}
\end{table*}

\begin{table*}[htp]
\centering
\caption{Post-Study Questionnaires: ranking of the system components to validate the competence of the system and determine the reliance on the system by all participants (All), domain experts, therapists (TPs), and novices (NVs) (i.e. health professionals and students)}
\label{tab:overall_questions}
\resizebox{\textwidth}{!}{%
\begin{tabular}{clll}\toprule
\textbf{Rank} & \multicolumn{1}{c}{\textbf{All}}                 & \multicolumn{1}{c}{\textbf{Therapists (TPs)}}      & \multicolumn{1}{c}{\textbf{Novices (NVs)}}             \\ \midrule
1 &
  Exploring confidence scores for task delegation (5.89) &
  { Exploring confidence scores for task delegation (6.0)} &
  A numerical confidence score of AI (5.89) \\
2    & A numerical confidence score of AI (5.81)        & A predicted score by AI (5.67)                     & Feature analysis (5.89)                                \\
3    & Feature analysis (5.74)                          & A numerical confidence score of AI (5.67)          & Exploring confidence scores for task delegation (5.86) \\
4    & A predicted score by AI (5.67)                   & Example-based explanation - Relevant Info (5.67)   & A distance-based confidence score of AI (5.78)         \\
5    & A distance-based confidence score of AI (5.67)   & Example-based explanation - Relevant Images (5.67) & A video of patient's exercises (5.72)                  \\
6    & Example-based explanation - Relevant Info (5.59) & Task delegation to AI (5.56)                       & A predicted score by AI (5.67)                         \\
7 &
  Example-based explanation - Relevant Images (5.56) &
  A distance-based confidence score of AI (5.44) &
  Example-based explanation - Relevant Info (5.56) \\
8    & Task delegation to AI (5.48)                     & Feature analysis (5.44)                            & Example-based explanation - Relevant Images (5.5)      \\
9    & {A video of patient's exercises (5.44)}   & {A video of patient's exercises (4.89)}     & {Task delegation to AI (5.44)}                  \\
10   & Intro \& Tutorials (4.93)                        & Intro \& Tutorials (4.78)                          & Intro \& Tutorials (5.0)   \\ \bottomrule
\end{tabular}%
}
\end{table*}

\begin{figure}[htp!]
\centering
\includegraphics[width=1.0\columnwidth]{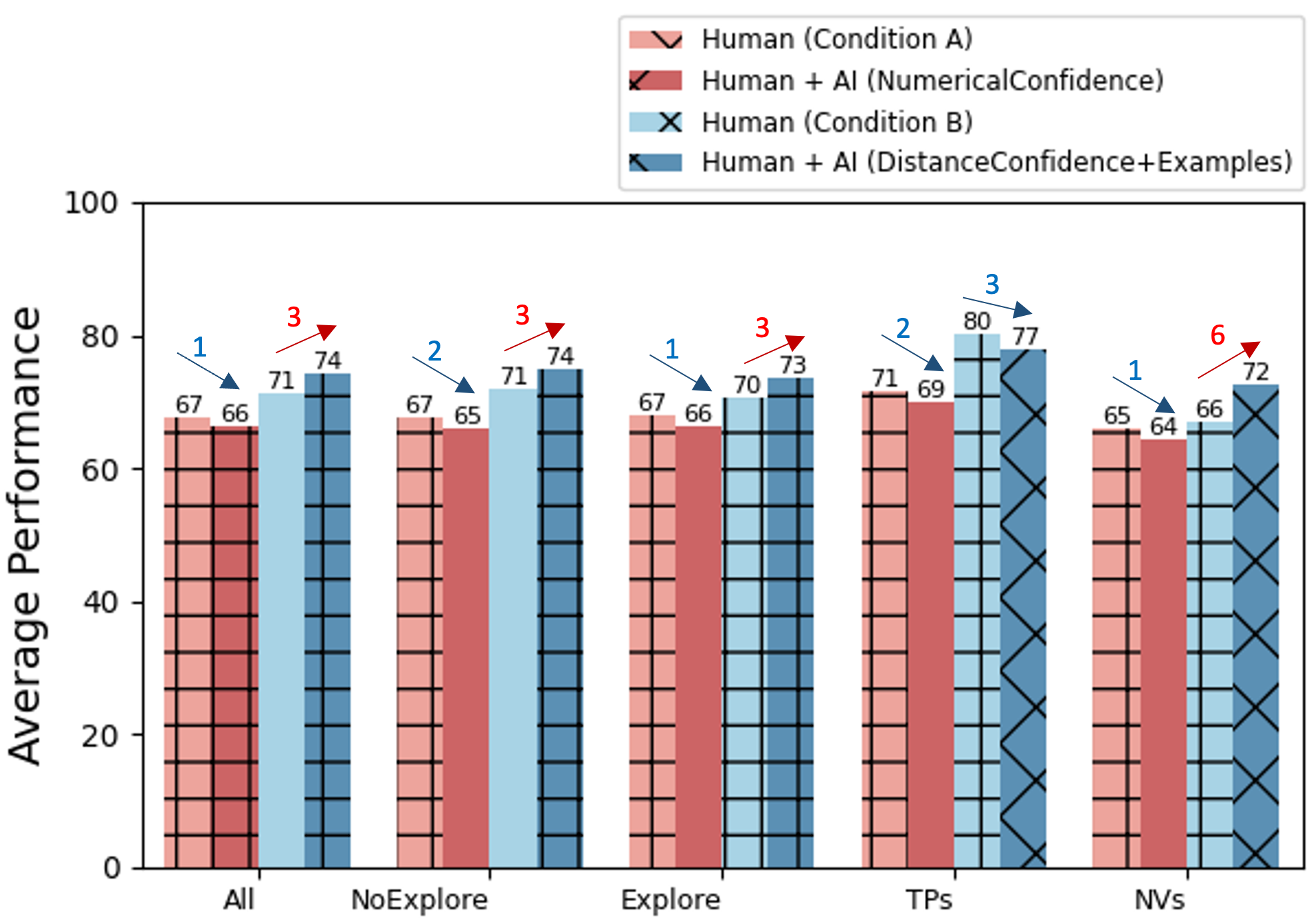}
\caption{Performance of AI-assisted decision-making on rehabilitation assessment tasks by all participants (All), participants without exploring a threshold of confidence scores (NoExp), participants with exploring a threshold of confidence scores (Exp), domain experts, therapists (TPs), and novices (NVs) (e.g. health professionals and students).}~\label{fig:results_performance}
\end{figure}

\end{document}